\pdfoutput=1
\documentclass[11pt]{article}
\PassOptionsToPackage{hyphens}{url}

\usepackage[utf8]{inputenc}
\usepackage[T1]{fontenc}

\usepackage{arxiv}

\usepackage{amsmath}
\usepackage{amssymb}
\usepackage{graphicx}
\usepackage{array}
\usepackage{booktabs}
\usepackage{listings}
\usepackage{subcaption}
\usepackage{enumitem}
\usepackage{natbib}
\usepackage{xcolor}
\usepackage{algorithm}
\usepackage{algpseudocode}

\usepackage{microtype}
\usepackage{xspace}
\usepackage{hyperref}
\usepackage{url}
\newcommand{\Order}{\textsc{Order}\xspace}
\newcommand{\Barrier}{\textsc{Barrier}\xspace}
\newcommand{\Durability}{\textsc{Durability}\xspace}
\newcommand{\Determinism}{\textsc{Determinism}\xspace}
\hypersetup{
  colorlinks=true,
  linkcolor=arxivaccent,
  citecolor=arxivaccent,
  urlcolor=arxivaccent,
  filecolor=arxivaccent,
  breaklinks=true,
}

\definecolor{codebg}{rgb}{0.95,0.95,0.97}
\definecolor{codegreen}{rgb}{0.0,0.5,0.0}
\definecolor{codegray}{rgb}{0.5,0.5,0.5}
\definecolor{codepurple}{rgb}{0.58,0,0.82}
\definecolor{codeblue}{rgb}{0.0,0.0,0.7}
\definecolor{alertbg}{rgb}{1.00,0.96,0.94}
\definecolor{alertred}{rgb}{0.70,0.10,0.10}

\lstdefinestyle{cstyle}{
    backgroundcolor=\color{codebg},
    basicstyle=\ttfamily\small,
    breaklines=true,
    captionpos=b,
    commentstyle=\color{codegreen},
    keywordstyle=\color{codeblue}\bfseries,
    numberstyle=\tiny\color{codegray},
    stringstyle=\color{codepurple},
    showstringspaces=false,
    numbers=left, numbersep=5pt, frame=single, rulecolor=\color{codegray}, tabsize=4,
}
\lstdefinestyle{shellstyle}{
    backgroundcolor=\color{codebg},
    basicstyle=\ttfamily\footnotesize,
    breaklines=true, language={}, showstringspaces=false,
    numbers=none, frame=single, rulecolor=\color{codegray}, tabsize=2,
}
\title{OneBarrier: What a Network Must Provide for\\Transparent Fault Tolerance to Be Free}

\author{
  Bojie Li \\
  Pine AI
}

\date{}
\runningtitle{OneBarrier: Transparent Fault Tolerance over a Total-Order Network}

\begin{document}
\maketitle

\begin{abstract}
Transparent fault tolerance---making an unmodified server binary survive
crashes---has been pursued for four decades without reaching production. Every attempt
paid three costs on the critical path: recording message arrival order for replay,
coordinating a consistent snapshot, and holding each reply until the state that produced
it was durable. This paper argues the costs are not intrinsic: they are the price of a
network that guarantees neither order nor delivery. We state four conditions under which
all three vanish. Three concern the network: \Order (messages are delivered in one
global sequence), \Barrier (delivery is confirmed by a commit barrier), and \Durability
(each message is replicated to backups before its barrier completes). The fourth,
\Determinism, falls to
the host: a user-space shim closes it for unmodified binaries at $2$--$10\%$
overhead---virtual time, virtualized randomness, and share-nothing sharding in place of
thread scheduling. OneBarrier realizes all four
conditions over an in-network total-order fabric (1Pipe) with microsecond round trips.
Fifteen unmodified applications---including Redis, Memcached, Nginx, Node.js, and a
multi-process PostgreSQL---recover byte-identically, and crash injection confirms
linearizable, exactly-once histories; the core protocols are machine-checked in TLA+. A
durable write placed inside the barrier adds $4.6\,\mu s$ to a request; the same write
placed after it adds three milliseconds. On a network that meets the conditions, fault
tolerance is a property, not a tax.
\end{abstract}

\begin{center}
\small
Code: \url{https://github.com/19PINE-AI/OneBarrier} \\[2pt]
Website: \url{https://01.me/research/OneBarrier}
\end{center}
\vspace{-0.2em}

\begin{center}
\includegraphics[width=0.98\linewidth]{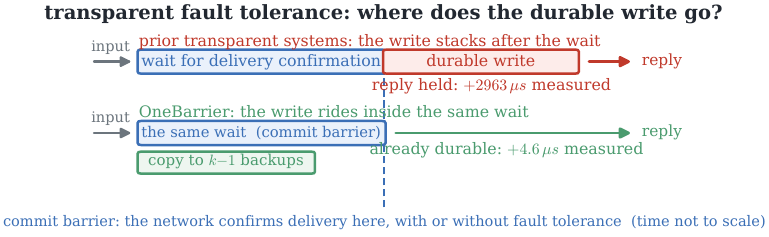}
\captionof{figure}{One request under transparent fault tolerance. Every prior
transparent system stacked the durable write after the delivery-confirmation
wait, holding the reply; OneBarrier places the copy to backups \emph{inside}
the commit barrier the network crosses anyway (\Durability), so the reply
leaves at the barrier. Measured on the real engine: $+4.6\,\mu s$ inside
versus $+2963\,\mu s$ after (\S\ref{sec:eval-overlap}).}
\label{fig:headline}
\end{center}
\vspace{-0.4em}


\section{Introduction}
\label{sec:intro}

A fault-tolerant service keeps running when a machine crashes. Most
services achieve this by being written for it: state is externalized into
a replicated store, or the application is restructured around a
replay-friendly framework. \emph{Transparent} fault tolerance requires no
such rewrite: take an unmodified server binary---a stock Redis, Nginx, or
PostgreSQL---and make it crash-recoverable from the outside, so that
clients lose no acknowledged work and never observe a repeated effect.

Transparent fault tolerance has been pursued for four decades and has
never reached production. The obstacle was not checkpointing but three
costs that every transparent system paid on the critical path of every
request~\citep{elnozahy2002survey}:
\begin{enumerate}[leftmargin=1.4em,itemsep=1pt,topsep=2pt]
\item \textbf{The order log.} Recovery re-executes (\emph{replays}) the
  failed server's inputs, which works only if they are re-applied in the
  original order. On an ordinary network that order is an accident of
  timing, so every message's arrival order must be recorded---the dominant
  overhead of deterministic-replay
  systems~\citep{mashtizadeh2017castor,ocallahan2017rr}.
\item \textbf{The coordinated snapshot.} A distributed checkpoint must be
  globally consistent: no node may record the receipt of a message that no
  node records sending. The classical
  solution~\citep{chandy1985snapshots} coordinates the cut with marker
  messages and records in-flight channels.
\item \textbf{The output hold.} A reply that has reached a client cannot
  be un-sent, so every externally visible output must be withheld until
  the state that produced it is durable---the \emph{output-commit}
  problem~\citep{strom1985optimistic}. This hold is what sank
  Remus~\citep{cully2008remus}: tens of milliseconds added to every reply.
\end{enumerate}
Faced with these costs, industry took the other road: rewrite the
application so that a purpose-built system holds the critical state, as in
durable-execution frameworks (Temporal~\citep{temporal},
DBOS~\citep{skiadopoulos2021dbos}, Restate~\citep{restate}) and
exactly-once stream processors (Flink~\citep{carbone2015flink}). That
works, one rewrite at a time, but it abandons the installed base whose
rewrite is precisely the cost transparency was meant to avoid.

\paragraph{Thesis.} The three costs are not intrinsic to fault tolerance.
They are the price of running over a network that promises nothing: no
ordering, no delivery confirmation, no redundancy. We state four explicit
conditions under which all three disappear. Three concern the network:
\begin{description}[leftmargin=1.4em,itemsep=1pt,topsep=2pt,font=\normalfont]
\item[\Order:] the network delivers all messages, at all
  receivers, in a single global order, identified by a timestamp on each
  message.
\item[\Barrier:] the network confirms
  delivery through a \emph{commit barrier}---a point in time at which a host
  knows that every message ordered up to timestamp $T$ has been delivered
  and none can be lost.
\item[\Durability:] each message is
  copied to backup nodes no later than its commit barrier.
\end{description}
Under \Order the order log vanishes: the network itself remembers the
order. Under \Order and \Barrier a snapshot needs no coordination: every
node independently cuts at the same timestamp. Under \Barrier and
\Durability the output hold costs nothing: the durability wait ends at the
same barrier the reply already waits for. Fault tolerance then adds no
round trip to any request---it is a byproduct of guarantees the network
provides for its own correctness (\S\ref{sec:conditions}).

The fourth condition falls to the host:
\begin{description}[leftmargin=1.4em,itemsep=1pt,topsep=2pt,font=\normalfont]
\item[\Determinism:] given the same ordered inputs,
  the server computes the same state and outputs. Real servers violate this
  locally---they read the clock, draw random numbers, and interleave
  threads.
\end{description}
The bulk of our experimental work shows that \Determinism is closable
\emph{from the outside}, for unmodified binaries, at $2$--$10\%$
overhead: an \texttt{LD\_PRELOAD} shim virtualizes time and randomness,
and share-nothing sharding replaces thread scheduling
(\S\ref{sec:libos}).

\paragraph{The system.} OneBarrier realizes the four conditions end to
end. \Order, \Barrier, and \Durability are supplied by
1Pipe~\citep{li2021onepipe}, a data-center communication layer that orders
messages inside the network switches and confirms delivery with an
aggregated commit barrier, at microsecond round trips over RDMA
(\S\ref{sec:fabric}). \Determinism
is supplied by our shim. The name records the central identity: the
output-commit hold and the network's reliable-delivery barrier are
\emph{one barrier}.

\paragraph{Evidence.} Fifteen unmodified applications recover
byte-identically---Redis, Memcached, Nginx, and Node.js by replay; a
multi-process PostgreSQL by whole-process checkpoint; ten more spanning
brokers, databases, a network function, a microservice, and
infrastructure daemons---and crash injection confirms linearizability and
exactly-once semantics (\S\ref{sec:eval}). On our engine, a durable write
placed within the commit barrier adds $4.6\,\mu s$ to a request; placed
after it, three milliseconds. A calibrated discrete-event model extends
the structure to 1Pipe's published microsecond operating point (we lack a
switch testbed), and the ordering and recovery protocols are
machine-checked in TLA+.

\paragraph{Contributions.}
\begin{itemize}[leftmargin=1.4em,itemsep=1pt,topsep=2pt]
\item The four conditions under which transparent fault tolerance adds no
  critical-path cost, with structural arguments for each cost's
  elimination and machine-checked protocol specifications
  (\S\ref{sec:conditions}, Appendix~\ref{app:tla}). The conditions double
  as an analysis of prior systems (\S\ref{sec:why}).
\item A user-space determinism shim that closes \Determinism for
  unmodified binaries---virtual time, virtualized randomness, share-nothing
  sharding---with a characterization of where the boundary lies: which
  servers can be replayed and which must fall back to checkpointing
  (\S\ref{sec:libos}).
\item An end-to-end system and evaluation: fifteen unmodified applications
  recovered byte-identically, correctness verified under crash injection,
  the ride-versus-stack structure measured, and cost quantified against
  same-engine reimplementations of the eliminated mechanisms
  (\S\ref{sec:system}, \S\ref{sec:eval}).
\end{itemize}

\section{Why Transparent Fault Tolerance Has Been Expensive}
\label{sec:why}

Every transparent system must guarantee one thing: at the moment any
output becomes externally visible, the state that produced it is
recoverable. The classical rollback-recovery
toolkit~\citep{elnozahy2002survey} offers three mechanisms toward that
guarantee---deterministic replay, consistent snapshots, output
commit---each carrying the corresponding cost of \S\ref{sec:intro}. The
three interlock: weakening one strengthens the demands on the others, so
prior systems could choose \emph{where} to pay but not \emph{whether}.

\begin{figure}[tb]
\centering
\includegraphics[width=0.62\linewidth]{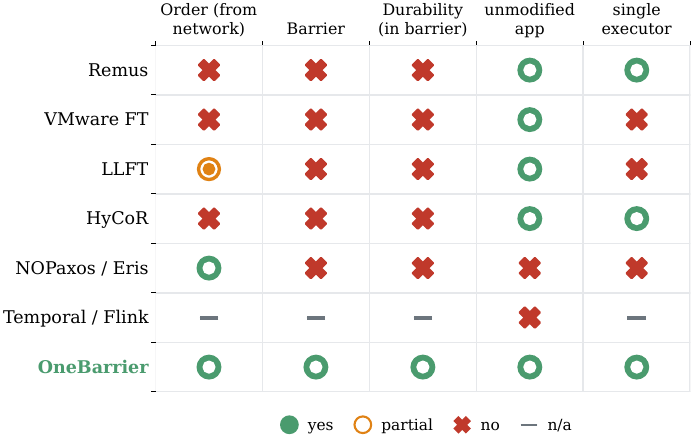}
\caption{Conditions satisfied by each system, and whether it serves
unmodified applications with a single executing replica.}
\label{fig:prior}
\end{figure}

Table~\ref{tab:prior} and Figure~\ref{fig:prior} organize the major
prior systems by which of our four conditions they satisfied. Read
this way, four decades of designs form a pattern: \emph{the condition a
system lacked is the cost it paid---or the guarantee it gave up.}

Remus~\citep{cully2008remus} satisfied none of the network conditions: it
ran whole virtual machines over ordinary Ethernet, abandoned replay
altogether (no order), checkpointed every few tens of milliseconds, and
buffered all output until the next checkpoint was durable---the output
hold in full. VMware FT~\citep{scales2010vmwareft} obtained order by brute
force, shipping every non-deterministic event to a lock-step backup; it
paid the order log as serialized, uniprocessor execution.
LLFT~\citep{zhao2013llft} came closest in spirit to \Order: a total order
imposed in \emph{host software} (leader-determined virtual
synchrony~\citep{birman1987isis}), which does eliminate the order log. But
the primary becomes a sequencer, every backup also executes every request
(semi-active replication), and, lacking \Barrier and \Durability, LLFT
declined the output hold rather than making it cheap: the primary replies
\emph{before} its backups know the ordering, which is safe only because
every peer must itself be an LLFT group holding the piggybacked ordering
information. The guarantee is surrendered at the boundary of that closed
world. HyCoR~\citep{zhou2021hycor} combined checkpoints with short replay
windows and logged message order within each window---the order log again.
On the other side of the ledger, systems that \emph{do} take order from
the network---NOPaxos~\citep{li2016nopaxos} and Eris~\citep{li2017eris},
which sequence messages in a programmable switch---deliberately stop at
ordering: their primitive is explicitly \emph{unreliable}, never
confirming, holding, or copying a packet, so hosts rebuild reliability and
durability with a quorum protocol, every replica logs and (at least
lazily) executes the full stream, and applications are written against the
protocol library. Derecho~\citep{jha2019derecho} brings ordered, reliable
group communication to RDMA clusters, again as a library for rewritten,
actively replicated applications. These systems satisfy \Order; none is
transparent, and none makes durability a byproduct of a delivery barrier.

\begin{table}[tb]\centering\small
\begin{tabular}{>{\raggedright\arraybackslash}p{0.16\linewidth}>{\raggedright\arraybackslash}p{0.27\linewidth}>{\raggedright\arraybackslash}p{0.22\linewidth}>{\raggedright\arraybackslash}p{0.24\linewidth}}
\toprule
system & where order comes from & where durability lives & cost paid on critical path \\
\midrule
Remus & none (no replay) & epoch checkpoint copy & output held tens of ms \\
VMware FT & event log, lock-step & backup VM in lock-step & serialized execution \\
LLFT & host-software total order & executing backups (in-memory) & host sequencer; $N\times$ execution; closed world \\
HyCoR & per-window order log & checkpoint + log & order log per message \\
NOPaxos / Eris & in-network order (unreliable) & host quorum of logs & app rewrite; $N\times$ (lazy) execution \\
Temporal / Flink & application rewrite & external store & the rewrite itself \\
\midrule
\textbf{OneBarrier} & in-network order (\Order) & in-barrier replicas (\Durability) & none beyond the barrier (\Barrier) \\
\bottomrule
\end{tabular}
\caption{Prior systems, organized by the conditions of \S\ref{sec:intro}.
The condition a system lacked is the cost it paid, or the guarantee it
gave up.}
\label{tab:prior}
\end{table}

No prior system satisfied \Order, \Barrier, and \Durability
simultaneously, and none combined them with a demonstration that
\Determinism is closable in practice for real, unmodified binaries. Those
two gaps are what this paper fills.

\section{Four Conditions That Make Fault Tolerance Free}
\label{sec:conditions}

\subsection{The conditions, precisely}
\label{sec:defs}

One global order of events through logical timestamps is due to
\citet{lamport1978time}, and replication has been built on ordered
delivery ever since: atomic
broadcast~\citep{birman1987isis,defago2004total}, and the state-machine
approach~\citep{schneider1990smr,lamport1998parttime}, in which replicas fed
identical inputs in identical order remain identical. Our conditions
package these classical ingredients as properties a \emph{network} can
provide to an \emph{unmodified} application:

\begin{description}[leftmargin=1.4em,itemsep=2pt,topsep=2pt,font=\normalfont]
\item[\Order:] Every message carries a timestamp drawn from one global
  order, and every receiver's messages are delivered in timestamp order.
\item[\Barrier:] For every timestamp $T$, each host eventually learns that
  \emph{all} messages with timestamps $\le T$ have been delivered and are no
  longer at risk of loss. We call the instant at which the host learns this
  the \emph{commit barrier for $T$}. (This is the delivery-confirmation phase of
  a two-phase commit~\citep{gray1978notes}, aggregated across the network.)
\item[\Durability:] When a message is delivered to its destination, a copy of
  it reaches $k-1$ backup nodes---at the same position in the global
  order---no later than that message's commit barrier.
\item[\Determinism:] The application is a deterministic function of its
  ordered input sequence: same inputs, same order $\Rightarrow$ same state
  and same outputs.
\end{description}

\begin{figure}[tb]
\centering
\includegraphics[width=0.68\linewidth]{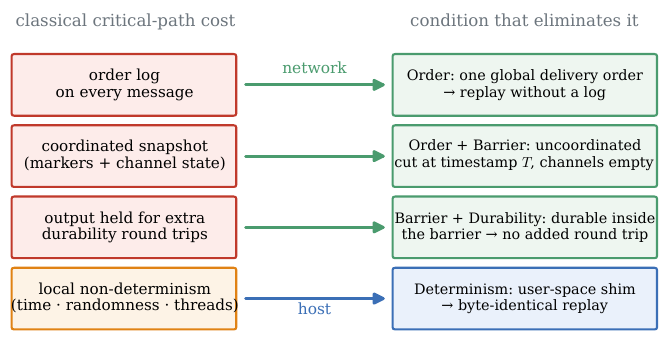}
\caption{The three classical costs of transparent fault tolerance, and the
condition that eliminates each. \Order, \Barrier, and \Durability are
properties of the network; \Determinism is manufactured for unmodified
binaries by a user-space shim.}
\label{fig:conditions}
\end{figure}

\Order, \Barrier, and \Durability are properties a network either has or
lacks; \Determinism belongs to the host, and \S\ref{sec:libos}
manufactures it for unmodified binaries. Figure~\ref{fig:conditions} maps
each classical cost to the condition that eliminates it.

\subsection{What each condition eliminates}
\label{sec:props}

\paragraph{\Order $+$ \Determinism: the order log vanishes.} Under
\Determinism, the server's state
after processing $n$ inputs is a function of those inputs and their order.
Under \Order, that order is the timestamp order, which every delivered
message carries. A recovering node therefore replays the saved inputs
sorted by their timestamps and reaches exactly the failed node's state---no
recorded order log, because the order was never an accident that needed
recording. This is state-machine
replication~\citep{schneider1990smr} with the network, rather than a
consensus protocol, as the sequencer.

\paragraph{\Order $+$ \Barrier: the snapshot needs no coordination.} To
checkpoint,
every node applies the same rule independently: process all inputs with
timestamps $\le T$, process none with timestamps $>T$, then save state. The
commit barrier for $T$ (\Barrier) tells the node when it has received
everything on its side of the line. Because sender and receiver classify
every message by the identical predicate---its timestamp---no message can
straddle the cut, so the cut is consistent \emph{and the channels are
empty}: there is no in-flight state to record. This is strictly simpler
than the classical marker protocol~\citep{chandy1985snapshots}, which
exists precisely to handle networks with no such global line. Cutting at a
synchronized timestamp appears in clock-synchronized
systems~\citep{corbett2012spanner,geng2018huygens}; \Order makes it exact,
because the timestamps \emph{are} the delivery order.

\begin{figure}[tb]
\centering
\includegraphics[width=0.58\linewidth]{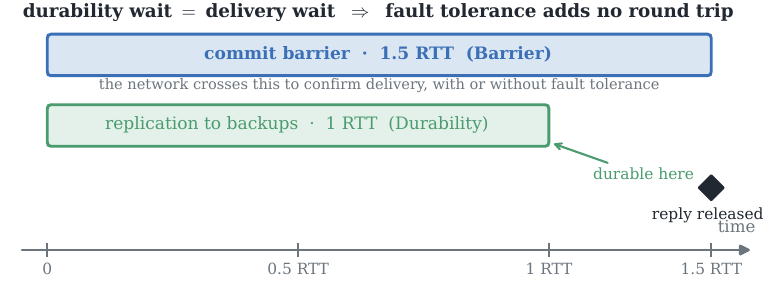}
\caption{\Durability inside \Barrier. Replication to backups (1 round
trip) completes before the commit barrier ($\sim$1.5 round trips) the
network crosses whether or not fault tolerance is enabled: the durability
wait and the delivery-confirmation wait end at the same barrier.}
\label{fig:barriertimeline}
\end{figure}

\paragraph{\Barrier $+$ \Durability: the output hold costs nothing.} Output
commit requires
a reply to wait until the state that produced it is durable. Under \Order
and \Determinism,
``the producing state'' is determined by the ordered input prefix, so the
requirement reduces to: \emph{the inputs up to this reply's cause must be
replicated}. \Durability says that replication completes by the commit
barrier; \Barrier says the reply already waits for that barrier, because
the network does not confirm delivery any earlier. The two waits---the
durability wait fault tolerance needs and the delivery-confirmation wait
the network imposes anyway---end at the same instant: not two stacked
costs but one barrier (Figure~\ref{fig:barriertimeline}). Fault tolerance
therefore adds no round trip at \emph{any} network speed; what speed sets
is the stakes. At microsecond round trips, the avoided round trip is the
difference between fault tolerance costing a multiple of request latency
and costing nothing. \S\ref{sec:eval-overlap} measures both the structure
and the magnitude.

We keep these structural arguments honest two ways: the ordering and
recovery protocols are machine-checked in TLA+ (Appendix~\ref{app:tla}),
and the third argument is tested by attempting to \emph{break} it
(\S\ref{sec:eval-overlap}).

\subsection{A network that meets Order, Barrier, and Durability}
\label{sec:fabric}

The conditions are not hypothetical. 1Pipe~\citep{li2021onepipe} is a
communication layer for data-center networks, demonstrated on commodity
RDMA network cards and a programmable switch~\citep{bosshart2014p4}, that
provides exactly \Order and \Barrier and contains the replication
primitive \Durability needs. We describe it from first principles.

\Order. Every sender stamps each message with a timestamp from
its local clock. Network switches, which see all traffic, continuously
compute and propagate a \emph{barrier}: the smallest timestamp that could
still be carried by any message in flight through them. A receiver holds
arriving messages briefly and releases them in timestamp order once the
barrier has passed them. The result is one global delivery order without
any central sequencer---the switch aggregation is the sequencer, scaling
with the network rather than with any single host.

\Barrier. On top of ordered best-effort delivery, 1Pipe's
reliable mode runs a two-phase commit: messages are first delivered and
acknowledged end-to-end (recovering any losses), and a second, aggregated
barrier round confirms to every participant that delivery up to timestamp
$T$ is complete---about half a round trip beyond the one round trip of
delivery, roughly $1.5$ round trips in all. On RDMA hardware, which reads
and writes remote memory without involving the remote CPU, a round trip is
$1$--$2\,\mu s$, and 1Pipe reports reliable delivery in roughly
$21\,\mu s$ at scale.

\Durability. Because delivery order is global, a sender can
\emph{scatter} copies of one message to several destinations at a single
position in the global order, in one round trip. Directing each input's
copies at $k-1$ backup nodes gives \Durability: the copies land within the
$1.5$-round-trip window the commit barrier already spans.

Nothing in \S\ref{sec:props} depends on these particular mechanisms or
speeds; any network meeting the definitions would do---a host-software
sequencer meets \Order, at the throughput cost we measure in
\S\ref{sec:eval-cost}. What the in-network realization contributes is that
the conditions hold \emph{below} the application at microsecond scale,
where the classical costs would otherwise dominate request latency.

\section{Manufacturing Determinism for Unmodified Binaries}
\label{sec:libos}

The network supplies the input order; the host must supply everything else
a binary needs to be a deterministic function of that order. The OneBarrier
determinism shim is three composable C libraries, loaded with
\texttt{LD\_PRELOAD} so that the dynamic linker resolves
the application's calls to standard functions (\texttt{recv},
\texttt{clock\_gettime}, \dots) against our interceptors before the C
library's. No kernel module, no recompilation, no application change.

\subsection{Which servers can be made deterministic: a fit test}
\label{sec:fittest}

Not every binary can be replayed, and the design itself says which can. A
server takes the \emph{replay} path if four questions answer yes: Is it
deterministic once input order and local non-determinism are fixed? Is it
share-nothing (or cleanly shardable into single-threaded instances)? Does
its input arrive through sockets the shim can intercept? Is each request's
output bounded, so there is a definite point at which the effect is
acknowledged? Servers that fail the share-nothing test---multi-process
databases with shared memory, arbitrary multi-threaded binaries---take a
second, fully general path: whole-process \emph{checkpointing}, which
needs no determinism at all (\S\ref{sec:criu}). The replay path is the one
on which fault tolerance is free, so the rest of this section closes its
remaining sources of local non-determinism: time, randomness, and
threads.

\subsection{Time: the boundary between two kinds of clock reads}
\label{sec:vclock}

\begin{figure}[tb]
\centering
\includegraphics[width=0.72\linewidth]{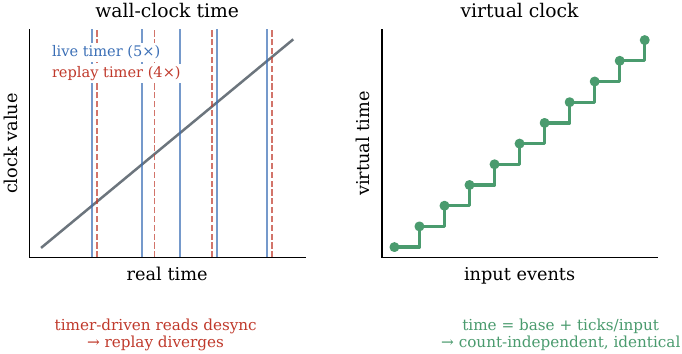}
\caption{The determinism boundary for time. Under the wall clock, a
background timer fires a different number of times in the original run and
the replay, so position-indexed replay desynchronizes (left). The virtual
clock advances only at input events, making every read a function of the
input prefix, identical across replay (right).}
\label{fig:boundary}
\end{figure}

Fixing time sounds simple---return recorded values---but a sharp boundary
separates the reads that a naive scheme handles from those that break it
(Figure~\ref{fig:boundary}). \emph{Request-driven} reads happen a fixed
number of times per request (a timestamp formatted into a reply); replaying
them by sequence position works. \emph{Timer-driven} reads come from
background timers that fire on wall-clock time---Redis's periodic
housekeeping (\texttt{serverCron}), Nginx's cached-time update. Between
the original run and the replay, real time elapses differently, so such a
timer fires a \emph{different number of times}, and position-indexed
replay falls out of step.

The shim's \emph{virtual clock} closes the boundary by making time a
function of the input prefix rather than of how often it is read. It
intercepts the C library's time surface (\texttt{gettimeofday},
\texttt{clock\_gettime}, \texttt{time}) and returns
$t=\mathrm{base}+\mathrm{ticks}$, where \texttt{ticks} advances only when
an input message arrives---by the \emph{real} inter-arrival gap, which is
logged during the original run and replayed from the log during recovery.
Two properties follow: reads are count-independent (a timer that fires
three or five times between two requests reads the same virtual time, so
timer-driven reads cannot desynchronize replay), and the clock stays
faithful to wall time (TTLs expire at true wall-clock moments, not on an
input-count schedule). Over the fabric, the logged gap is derived from the
message's own global timestamp, already durable under \Durability---the
time log and the input log are the same log (code in
Appendix~\ref{app:libos}). The sharpest demonstration is Nginx: the
\texttt{Date} header it formats deep inside its own time cache is
byte-identical between the live run and a recovery seconds later.

\subsection{Randomness: a per-consumer boundary}
\label{sec:rng}

Randomness reaches a server by more paths than the C library, and each path
must be pinned where it actually flows. Node.js's V8 engine, OpenSSL, and
\texttt{arc4random} issue the raw \texttt{getrandom} system call,
bypassing any intercepted library function; Redis seeds its hash tables by
reading \texttt{/dev/urandom} through \texttt{fopen}, whose internal calls
likewise bypass the public symbols. The shim therefore intercepts at the
system-call level: a seccomp filter (a kernel mechanism that lets a
designated user-space supervisor handle chosen system calls) traps
\texttt{getrandom} and fills the buffer from a seeded deterministic
stream, and a private mount namespace redirects \texttt{/dev/urandom} to a
deterministic file. This pins every consumer routing through those
interfaces---OpenSSL's generator, and Redis's hash seed, whose dependent
command orderings (\texttt{SPOP}, \texttt{SRANDMEMBER}) recover
identically where an unpinned control differs entirely. One consumer marks
the limit: V8 seeds \texttt{Math.random} from the CPU's \texttt{RDRAND}
instruction---no system call at all---so we pin it at the layer that owns
it, a recorded \texttt{--random-seed} launch flag (\S\ref{sec:limits}).
The lesson: the randomness boundary is per-\emph{consumer}, not
per-interface.

\subsection{Threads: sharding, not scheduling}
\label{sec:threads}

\begin{figure}[tb]
\centering
\begin{subfigure}{0.43\linewidth}\centering
\includegraphics[width=\linewidth]{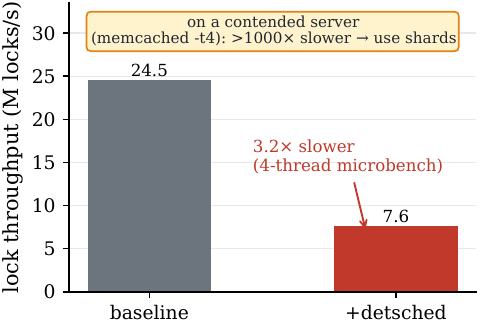}
\caption{Deterministic scheduling serializes critical sections.}
\label{fig:detsched}
\end{subfigure}\hfill
\begin{subfigure}{0.43\linewidth}\centering
\includegraphics[width=\linewidth]{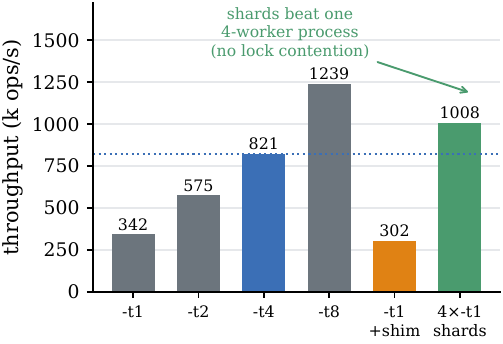}
\caption{Share-nothing shards beat one multi-worker process.}
\label{fig:sharding}
\end{subfigure}
\caption{Meeting \Determinism for multi-threaded servers. (a) A Kendo-style
deterministic scheduler collapses throughput on a contended server. (b) The
practical path is share-nothing sharding: four single-thread Memcached
shards exceed one four-worker process while staying deterministic (each
shard pays $\sim$10\% shim overhead; the aggregate is sub-linear because
shards and the load generator share one host's cores and NIC).}
\label{fig:threads}
\end{figure}

In a multi-threaded server the kernel scheduler decides the order in which
threads enter critical sections, so shared state evolves
non-deterministically. Deterministic-scheduling systems in the style of
Kendo~\citep{olszewski2009kendo,bergan2010coredet} fix this by admitting
threads to locks in logical-clock order, but the fix serializes critical
sections; on a contended server the throughput collapse is severe
(Figure~\ref{fig:threads}a), which is the recognized wall of that
literature and of execute-then-verify
replication~\citep{kapritsos2012eve,guo2014rex}. We adopt the answer
high-performance servers already practice:
\emph{share-nothing sharding}. Run $N$ single-threaded instances, each
deterministic by construction, each its own replay unit; there is no lock
contention to serialize. Four single-threaded Memcached shards out-throughput
one four-worker Memcached while remaining deterministic
(Figure~\ref{fig:threads}b). The deterministic scheduler remains a fallback
for genuinely shared mutable state. One residual: single-threaded servers
still spawn timer-driven maintenance threads (Memcached's LRU maintainer);
the shim disables these so a shard's state evolves purely from the request
stream and the virtual clock.

\subsection{Bounding replay: checkpoints, and the general path}
\label{sec:criu}

Replaying from process start costs time proportional to the log; a
checkpoint bounds recovery to the tail after it. For a single process an
application-native snapshot (such as a Redis RDB file) plus tail replay
suffices. The general mechanism, and the fallback path for binaries that
fail the fit test, is CRIU~\citep{criu}, a Linux facility that checkpoints a
whole process tree---memory, file descriptors, threads, and, in our use,
the shim's virtual clock---so a restore needs no replay of pre-checkpoint
history (in the user-space checkpointing lineage of
DMTCP~\citep{ansel2009dmtcp} and BLCR~\citep{hargrove2006blcr}). We verify
the hardest case in this class: a multi-process PostgreSQL (postmaster plus
background workers, shared memory segments, write-ahead log)
checkpoint-restores with byte-identical data and a live server
(\S\ref{sec:eval-recovery}). PostgreSQL exercises the checkpoint path
\emph{only}: its cross-process non-determinism (the order backends touch
shared buffers) is removed by neither sharding nor a per-process virtual
clock, so it lives in the checkpoint-only regime that Remus and HyCoR
also occupy.

\subsection{Output suppression: exactly-once across recovery}
\label{sec:suppress}

During replay the server re-computes---and tries to re-send---outputs it
already sent before the crash. The shim suppresses duplicates with a
per-client high-water mark: each output carries a per-client sequence
number, and an output at or below the client's last acknowledged sequence
is dropped. This is the exactly-once RPC discipline of
RIFL~\citep{lee2015rifl} applied to recovery. The high-water-mark table is
itself part of the state: it is captured in every timestamp-$T$ snapshot
and rebuilt by the same replay that rebuilds the application, so
suppression survives a crash that loses all volatile state. The resulting
invariants---every acknowledged write survives, no effect is externalized
twice---are machine-checked in TLA+ (Appendix~\ref{app:tla}). Suppression
is server-side; end-to-end exactly-once additionally requires the
\emph{client} to present a stable identity on reconnection---a fabric
endpoint has one by construction, an anonymous external TCP client does
not (\S\ref{sec:limits}).

\section{The OneBarrier System}
\label{sec:system}

\subsection{Architecture and protocol}

\begin{figure}[tb]
\centering
\includegraphics[width=0.74\linewidth]{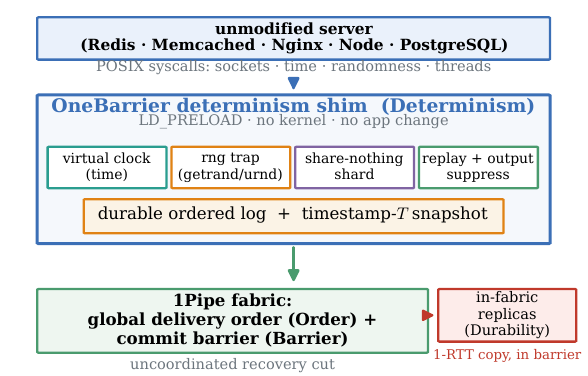}
\caption{OneBarrier. An unmodified server runs over the determinism shim
(\Determinism), which routes its sockets onto a fabric providing global
order (\Order), a delivery commit barrier (\Barrier), and in-barrier
replication of every input (\Durability). Replies are released at the
barrier---the same barrier the fabric crosses to confirm delivery.}
\label{fig:arch}
\end{figure}

Figure~\ref{fig:arch} assembles the pieces---an unmodified share-nothing
server over the determinism shim, the fabric supplying order, barrier,
and in-barrier replication---and Algorithm~\ref{alg:ft} gives the
per-replica protocol.

On the live path, the engine applies fabric-delivered inputs in timestamp
order, appends each to a durable ordered log, scatters it to the backups
(one round trip, within the barrier), and releases an output only when the
commit barrier passes the timestamp of the input that produced it. When the
barrier passes a snapshot point $T$, the engine quiesces---finishes inputs
$\le T$, defers inputs $>T$---and checkpoints; by \S\ref{sec:props} this
cut is consistent with no coordination. On recovery, a replacement loads
the latest snapshot, replays the log suffix in timestamp order (no order
log), fetches from a survivor any prefix it is missing, and resumes live
delivery. Replay runs the same transition function with external effects
disabled; the high-water marks drop every output the suffix re-derives, so
recovery is exactly-once.

\begin{algorithm}[tb]
\small
\caption{OneBarrier live and recovery paths (per replica).}
\label{alg:ft}
\begin{algorithmic}[1]
\Procedure{OnDeliver}{$(\mathit{ts}, \mathit{op})$, \textit{live}}\Comment{fabric delivers in \texttt{ts} order (\Order)}
  \State $\mathit{log}.\textsc{append}(\mathit{ts}, \mathit{op})$
  \If{\textit{live}}
     \State scatter input to $k-1$ backups at position \texttt{ts}\Comment{1 RTT, inside the barrier (\Durability)}
  \EndIf
  \State $\mathit{out} \gets \textsc{Apply}(\mathit{op})$\Comment{deterministic transition (\Determinism)}
  \If{$\mathit{out}.\mathit{seq} > \mathit{hwm}[\mathit{out}.\mathit{client}]$}\Comment{not yet externalized}
     \State $\mathit{hwm}[\mathit{out}.\mathit{client}] \gets \mathit{out}.\mathit{seq}$
     \If{\textit{live}} release $\mathit{out}$ at the commit barrier for \texttt{ts}\Comment{output commit $=$ \Barrier}
     \EndIf
  \EndIf\Comment{on replay, duplicate outputs are dropped: exactly-once}
\EndProcedure
\Procedure{OnBarrier}{$T$}\Comment{barrier passes snapshot point $T$}
  \State finish inputs $\le T$; defer inputs $> T$; \textsc{Checkpoint}($T$)\Comment{uncoordinated consistent cut}
\EndProcedure
\Procedure{Recover}{}
  \State \textsc{Load}(latest snapshot)\Comment{restores state \emph{and} $\mathit{hwm}$}
  \State fetch any missing prefix from a survivor, up to the live cut
  \State \textbf{for} each $(\mathit{ts}, \mathit{op})$ in log suffix, in \texttt{ts} order: \textsc{OnDeliver}($(\mathit{ts},\mathit{op})$, \textit{live}$=$\textsc{false})
  \State resume live delivery past the recovered cut
\EndProcedure
\end{algorithmic}
\end{algorithm}

\subsection{Durability, membership, and recovery under load}
\label{sec:membership}

The configuration is $k$ replicas of the log; an input is durable once any
survivor holds it, giving fail-stop tolerance of $f<k$ simultaneous
crashes with in-memory copies. Membership rides on the fabric: its failure
detector excises a dead peer within tens of microseconds and the survivors'
commit barrier resumes over the reduced group. Because OneBarrier is
\emph{passive}---one replica executes, the backups only log---a primary
failure opens a recovery window of unavailability; automated primary
promotion---a view change, which fundamentally requires
consensus~\citep{fischer1985impossibility,lamport1998parttime}---is
left to a production layer (\S\ref{sec:limits}). A recovering replica
converges only while replay outruns the live stream; the fabric's
barrier hold doubles as backpressure on senders, roughly doubling the live
load a recovery can outrun (Appendix~\ref{app:model}).

\subsection{Implementation and machine-checked protocols}
\label{sec:impl}

The engine is written in Rust over the reliable-delivery layer
of an open 1Pipe reproduction, implementing the ordered durable log,
timestamp-$T$ snapshots, output suppression, and crash recovery; the test
servers (speaking the Redis, Memcached, and HTTP protocols, plus a
transactional store and a publish/subscribe log), fault injectors, and
correctness checkers ship as
runnable binaries, and each quantitative claim is guarded by a named test.
The determinism shim is the three C libraries of \S\ref{sec:libos}. We
specify the two correctness-critical protocols in TLA+ and model-check
them (Appendix~\ref{app:tla}): the fabric's total-order delivery, and the
engine's exactly-once and no-lost-acknowledged-write invariants under
arbitrary crash/recover interleavings. The models verify the protocol
abstractions; the Rust and C implementations are validated separately by
the crash-injection and linearizability checks of \S\ref{sec:eval-correct}.
Every result in \S\ref{sec:eval} reproduces with a single command in the
open-source artifact.

\section{Evaluation}
\label{sec:eval}

Four questions, one per subsection:
\textbf{Q1}~Can \Determinism really be manufactured for unmodified
binaries (\S\ref{sec:eval-recovery})?
\textbf{Q2}~Is the recovered service \emph{correct}, not merely identical
(\S\ref{sec:eval-correct})?
\textbf{Q3}~Does the durable write ride the commit barrier or stack on top
of it---the load-bearing claim of \S\ref{sec:props}
(\S\ref{sec:eval-overlap})?
\textbf{Q4}~What does it all cost (\S\ref{sec:eval-cost})?

\emph{Evidence tiers, stated once.} Q1, Q2, and Q4 are measured on real,
unmodified binaries or on the real engine. For Q3, the ride-versus-stack
\emph{structure} is measured on the real engine over a loopback network;
its \emph{magnitude at the microsecond operating point} comes from a
discrete-event model calibrated against that engine, because we lack an
RDMA/programmable-switch testbed. We flag the modeled numbers where they
appear and nowhere claim them as measurements.

\subsection{Q1: Byte-identical recovery of fifteen unmodified applications}
\label{sec:eval-recovery}

\begin{figure}[tb]
\centering
\begin{subfigure}{0.44\linewidth}\centering
\includegraphics[width=\linewidth]{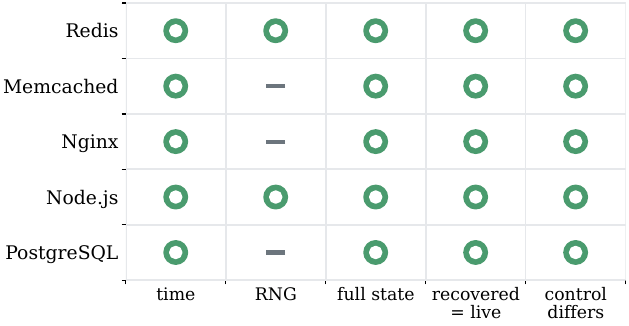}
\caption{Determinism coverage per server.}
\label{fig:recmatrix}
\end{subfigure}\hfill
\begin{subfigure}{0.33\linewidth}\centering
\includegraphics[width=\linewidth]{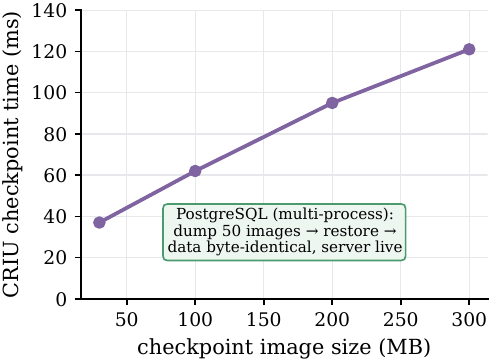}
\caption{CRIU checkpoint cost.}
\label{fig:criu}
\end{subfigure}
\caption{(a) Time, randomness, and full-state recovery for the five core
servers: recovered state equals live state, and a no-shim control differs
(a dash marks a probe the server does not exercise).
(b) Whole-process checkpoint time versus image size; a multi-process
PostgreSQL checkpoint-restores with byte-identical data.}
\label{fig:recovery}
\end{figure}

The test of transparency is blunt: kill an unmodified server with
\texttt{kill -9}, wait a real wall-clock gap, recover it, and demand that
its state match the original byte for byte. Four servers take the replay
path (record under the shim; crash; replay on a fresh instance):
\textbf{Redis} (TTL expirations and hash-seed-dependent command orders),
\textbf{Memcached} (LRU eviction order), \textbf{Nginx} (the
\texttt{Date} header from its internal time cache), and \textbf{Node.js}
(session identifiers derived from \texttt{Math.random()} and
\texttt{Date.now()}). \textbf{PostgreSQL}, shared-everything, takes the
CRIU path: the multi-process tree checkpoint-restores with byte-identical
on-disk data and a live server. In every case a control run without the
shim diverges---the determinism is manufactured, not incidental
(Figure~\ref{fig:recovery}). Robustness is not a lucky draw: across 25
independent record/crash/replay trials of Redis and 8 of Node.js, recovery
was byte-identical in every trial and the control differed in every
trial.

\paragraph{Ten further applications.} To test the fit test of
\S\ref{sec:fittest} beyond the servers it was designed on, we routed ten
further applications through it, spanning five
classes: message brokers (Redis Streams, a Kafka-style partitioned log,
Mosquitto MQTT), databases (SQLite under an unmodified Python process;
MySQL/MariaDB via CRIU), a stateful network function (a Click-style flow
table and load balancer~\citep{li2016clicknp,sherry2015ftmb}), a stateful
order/checkout microservice, and infrastructure daemons (dnsmasq,
lighttpd, HAProxy). Each recovers byte-identically along the path the fit
test assigns (Figure~\ref{fig:extresults}; per-class detail in
Appendix~\ref{app:ext}). Three results sharpen the boundary. A stock Redis
stream broker recovers its auto-generated entry IDs and consumer offsets
exactly---the guarantee exactly-once stream processors are rewritten to
provide. An unmodified SQLite-backed process recovers by deterministic
\emph{replay}, placing a SQL database on the free path and locating the
real boundary at shared-everything concurrency, not at ``databases.'' And
the determinism unit proved to be the \emph{message}, not the connection:
UDP servers (dnsmasq), not TCP ones, needed the small extension---the
opposite of the usual assumption.

\begin{figure}[tb]
\centering
\begin{subfigure}{0.45\linewidth}\centering
\includegraphics[width=\linewidth]{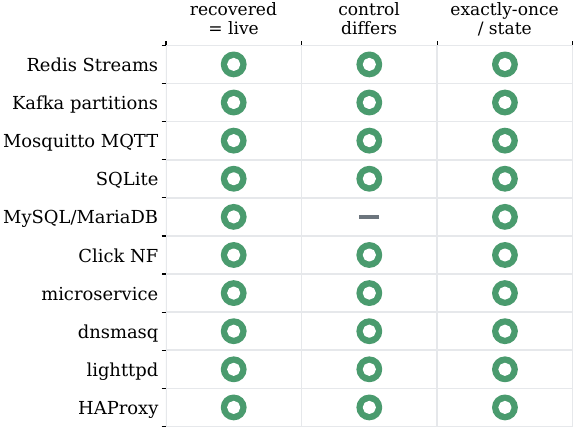}
\caption{Recovery holds for every extension app.}
\label{fig:extmatrix}
\end{subfigure}\hfill
\begin{subfigure}{0.53\linewidth}\centering
\includegraphics[width=\linewidth]{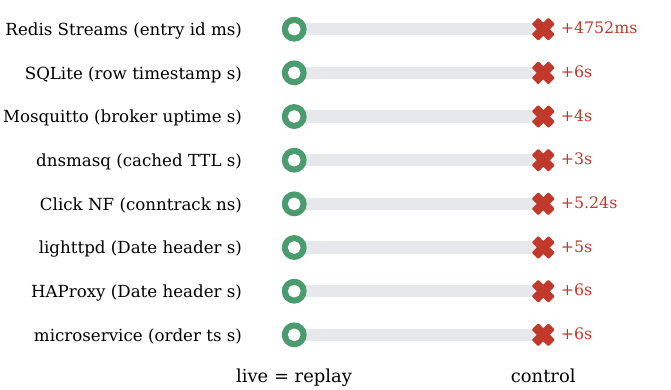}
\caption{Recovered replay $=$ live; control drifts.}
\label{fig:extdet}
\end{subfigure}
\caption{The ten extension applications across five classes. (a) Each
recovers byte-identically with the expected control-differs witness (a
dash marks the checkpoint path, which has no replay-versus-control axis).
(b) For the applications with a time- or randomness-derived probe,
recovery reproduces it exactly, while a no-shim control drifts by the
labeled amount.}
\label{fig:extresults}
\end{figure}

\subsection{Q2: Correctness under crash injection}
\label{sec:eval-correct}

Byte-identity shows the recovered state equals the recorded one; it does
not show that state was ever consistent. Two checks close the gap, both on
an unmodified Redis crashed mid-load under eight concurrent clients.
\emph{Survival at scale}: every acknowledged write reappears after
recovery with its exact value---$191{,}073$ acknowledged, $0$ lost, $0$
torn---with operations still in flight at the crash excluded, as output
commit prescribes. \emph{A genuine linearizability verdict}: a from-scratch
Wing--Gong checker~\citep{herlihy1990linearizability} searches the
real-time-overlapping history of a contended register, including the
post-recovery read, and certifies it linearizable. The two are
complementary: survival scales to hundreds of thousands of operations; the
linearizability search is exponential in concurrency, so its verdict is
rendered on a $33$-operation history.

\subsection{Q3: Does durability ride the barrier, or stack on it?}
\label{sec:eval-overlap}

Section~\ref{sec:props} claims that under \Barrier and \Durability the
durable
write adds no round trip because it completes inside a barrier the network
crosses anyway. The falsifiable form of the claim is comparative: put the
durable write \emph{inside} the barrier (\Durability holds) and its
marginal cost should be microseconds of copying, not a round trip; put it
\emph{after} the barrier (\Durability broken, as with a synchronous disk
write) and the classical output-commit cost should reappear in full. We
run both on the real engine over a loopback network
(Table~\ref{tab:rq2}).

\begin{table}[tb]\centering\small
\begin{tabular}{lrrr}
\toprule
durability placement & delivery p50 & marginal durability cost & commit p50 \\
\midrule
inside the barrier (\Durability holds) & $2014\,\mu s$ & $\boldsymbol{+4.59\,\mu s}$ & $2018\,\mu s$ \\
after the barrier (fsync)    & $3042\,\mu s$ & $\boldsymbol{+2963\,\mu s}$ & $6016\,\mu s$ \\
\bottomrule
\end{tabular}
\caption{Ride versus stack, measured on the real engine (loopback network,
p50). Replication inside the barrier adds microseconds of copying to the
commit path; a serial write after the barrier adds milliseconds and
saturates the executor. The marginal durability cost is the p50 of the
\emph{paired per-request} difference against the no-durability baseline,
so it does not equal a difference of the column medians.}
\label{tab:rq2}
\end{table}

\begin{figure}[tb]
\centering
\includegraphics[width=0.88\linewidth]{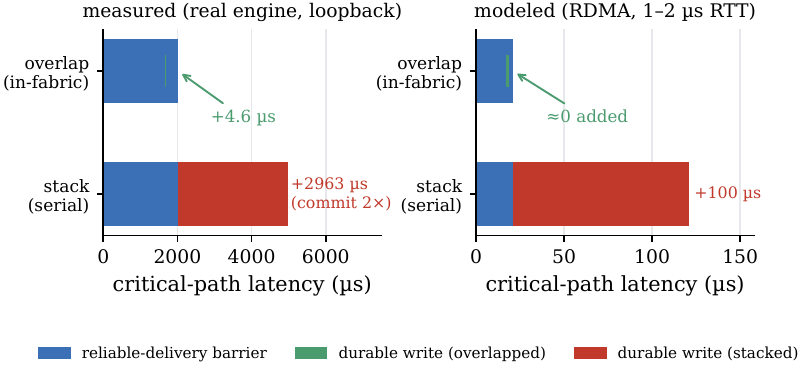}
\caption{Ride versus stack. Top bars: the durable write (green) completes
under the commit barrier the network crosses regardless, adding
microseconds of copying. Bottom bars: placed after the barrier, the same
write stacks (red). Left: measured on the real engine over loopback
($+4.6\,\mu s$ versus $+2963\,\mu s$). Right: the same structure at the
modeled RDMA operating point, where a $\sim$$1.5\,\mu s$ replication write
hides inside a $\sim$$21\,\mu s$ barrier while serial storage
($\sim$$100\,\mu s$) stacks.}
\label{fig:overlap}
\end{figure}

The structure is unambiguous: in-barrier durability costs microseconds of
copying; post-barrier durability costs milliseconds, doubles commit
latency, and reproduces the failure mode that sank Remus
(Figure~\ref{fig:overlap}). The loopback delivery path is itself
millisecond-scale, so dividing $4.6\,\mu s$ by it yields a flattering
fraction ($0.2\%$) we do not lean on; what the experiment establishes is
the \emph{structure}---ride versus stack.

\begin{figure}[tb]
\centering
\includegraphics[width=0.82\linewidth]{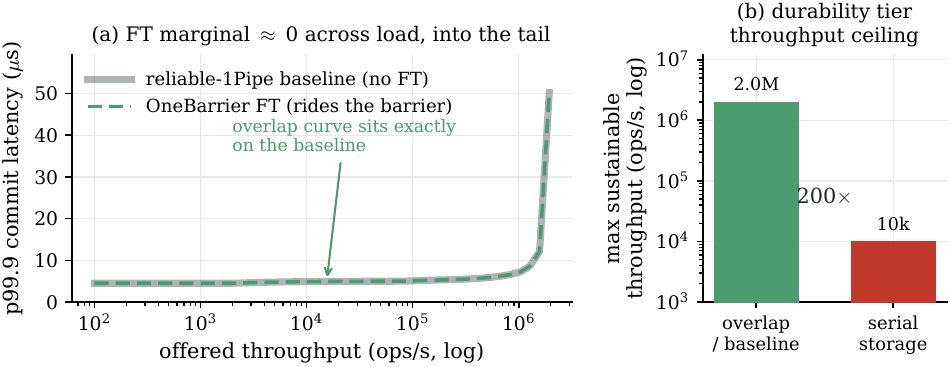}
\caption{Load sweep at the modeled RDMA operating point. (a) p99.9 commit
latency versus offered load: the fault-tolerant tier (durability inside
the barrier) is coincident with the no-FT baseline across the entire load
range. (b) The serial-durability tier's throughput ceiling is set by its
$\sim$$100\,\mu s$ post-barrier write, $200\times$ below the apply-bound
ceiling---the classical output-commit collapse, quantified across load.}
\label{fig:loadsweep}
\end{figure}

At the microsecond operating point---the 1Pipe pipeline at
its published parameters ($1$--$2\,\mu s$ round trip,
$\sim$$21\,\mu s$ reliable-delivery barrier, $\sim$$1.5\,\mu s$
replication write; Appendix~\ref{app:model})---if the write
\emph{stacked}, it would add $\sim$$1.5\,\mu s$ to a $21\,\mu s$ path
($7\%$) in the best case and $\sim$$100\,\mu s$ ($5\times$) with serial
storage---so the operating point is exactly where riding versus stacking
changes the verdict on deployability. In the model, the fault-tolerant
tier's commit-latency distribution is indistinguishable from the
no-fault-tolerance baseline at every offered load, into the p99.9 tail
(Figure~\ref{fig:loadsweep}), while the serial tier's throughput ceiling
sits two orders of magnitude below.

The model is not free-floating: fed the
engine's measured loopback latencies, it reproduces the in-barrier commit
latency to $\sim$$0.1\%$ and predicts the serial tier's saturation knee,
which the real engine reproduces; real RDMA transfers over a software RoCE
stack corroborate the one-sided-write latency it assumes, though only
switch hardware would settle the full round-trip pipeline
(Appendix~\ref{app:model}).

\subsection{Q4: Costs}
\label{sec:eval-cost}

\begin{figure}[tb]
\centering
\begin{subfigure}{0.43\linewidth}\centering
\includegraphics[width=\linewidth]{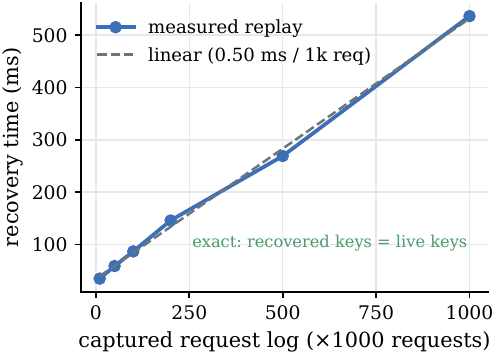}
\caption{Recovery time vs.\ log length.}
\label{fig:rectime}
\end{subfigure}\hfill
\begin{subfigure}{0.43\linewidth}\centering
\includegraphics[width=\linewidth]{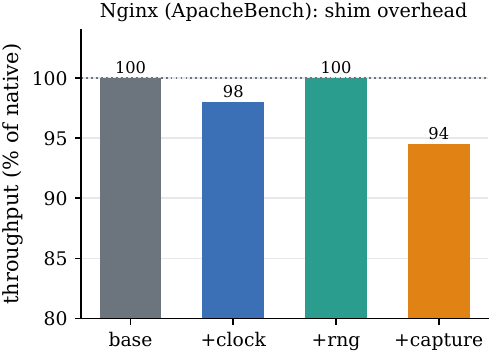}
\caption{Steady-state shim overhead (Nginx).}
\label{fig:overhead}
\end{subfigure}
\caption{Costs on real binaries. (a) Recovery time is affine in the
replayed log ($\sim$$30$\,ms floor $+$ $\sim$$0.5$\,ms per 1000 requests);
checkpoints bound it. (b) Steady-state overhead on Nginx: time
$\sim$$2\%$, randomness within noise, capture $<$$5\%$.}
\label{fig:costs}
\end{figure}

\paragraph{Steady-state overhead.} The virtualization a server actually
exercises is cheap because each interceptor is a thin wrapper over a call
the server already makes: on Nginx under ApacheBench, time virtualization
costs $\sim$$2\%$, randomness virtualization is within noise, and input
capture costs $<$$5\%$ (Figure~\ref{fig:costs}b). The one costly mechanism
is deterministic scheduling, whose serialization is fundamental
(\S\ref{sec:threads})---which is why the production path is sharding.

\begin{figure}[tb]
\centering
\begin{subfigure}{0.43\linewidth}\centering
\includegraphics[width=\linewidth]{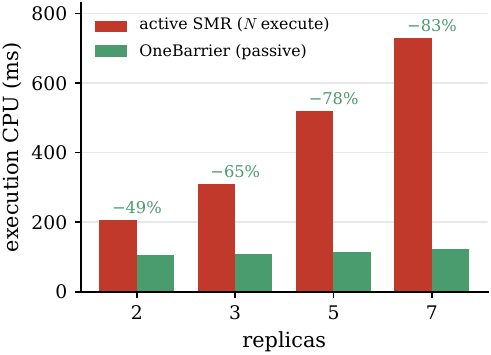}
\caption{Passive vs.\ active replication: execution CPU.}
\label{fig:passive}
\end{subfigure}\hfill
\begin{subfigure}{0.43\linewidth}\centering
\includegraphics[width=\linewidth]{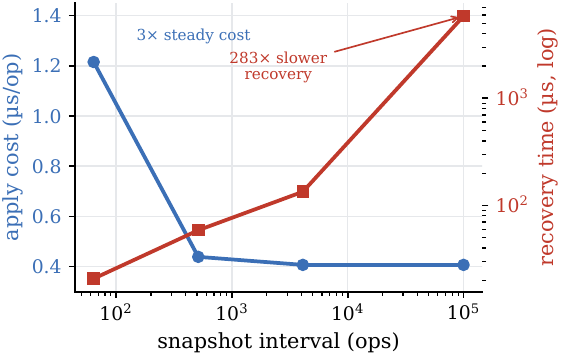}
\caption{Snapshot-interval tradeoff.}
\label{fig:interval}
\end{subfigure}
\caption{Resource cost. (a) Passive replication holds execution CPU near
$1\times$ ($106$--$123$\,ms) as replicas grow, where active replication
scales linearly ($206$--$730$\,ms): a $49$--$83\%$ saving. (b) A small
snapshot interval triples steady apply cost but restores the engine's
in-memory state in $\sim$$20\,\mu s$; a large interval is the reverse
($283\times$ slower restore).}
\label{fig:resource}
\end{figure}

\paragraph{Recovery time.} Recovery is affine in the replayed log length
above a $\sim$$30$\,ms restore floor---about $0.5$\,ms per thousand
requests (Figure~\ref{fig:costs}a)---and a checkpoint caps the replayed
suffix, so downtime is a tunable parameter: a checkpoint every $10^5$
requests holds recovery under $100$\,ms, versus the seconds-to-minutes of
detect-and-restart failover. The snapshot interval trades steady-state
apply cost against recovery time in the expected way
(Figure~\ref{fig:resource}b).

\begin{figure}[tb]
\centering
\includegraphics[width=0.52\linewidth]{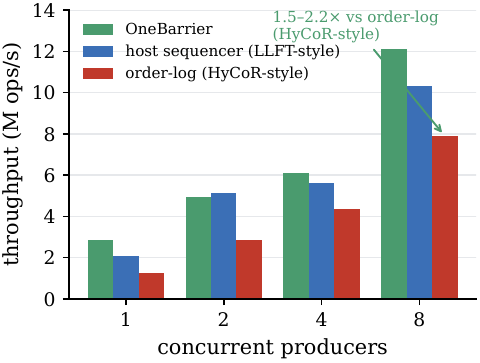}
\caption{Throughput against same-engine reimplementations of the
mechanisms the conditions eliminate: the per-input order log (\Order
removes it) and the host-software sequencer (in-network ordering removes
it). OneBarrier beats the order-log tier by $1.5$--$2.2\times$.}
\label{fig:competitors}
\end{figure}

\paragraph{Against the mechanisms the conditions eliminate.} To isolate
what \Order and \Durability save, we reimplemented the two mechanisms they
replace \emph{inside the same engine}---a HyCoR-style per-input order log
and an LLFT-style host-software sequencer---so everything else is held
equal (these are mechanism baselines, not the released systems). Removing
the order log yields $1.5$--$2.2\times$ higher throughput and strictly
fewer durable bytes (Figure~\ref{fig:competitors}); the host sequencer is
masked by apply work at low producer counts and becomes the bottleneck
under contention, where an in-network sequencer scales with the network
(1Pipe reports $2$--$20\times$ on hardware~\citep{li2021onepipe}). And
because OneBarrier is passive, backups only log: aggregate execution CPU
stays near a single replica's as the replication factor grows, a
$49$--$83\%$ saving over active replication in which every replica
executes (Figure~\ref{fig:resource}a).

\section{Limitations}
\label{sec:limits}

\textbf{No switch/RDMA testbed.} The ride-versus-stack structure is
measured on the real engine; its magnitude at the microsecond operating
point rests on a calibrated model and published 1Pipe measurements
(\S\ref{sec:eval-overlap}, Appendix~\ref{app:model}). This is the one
major claim not measured on real artifacts.
\textbf{Durability model.} Durability is in-memory replication tolerating
$f<k$ fail-stop crashes, not persistence across correlated power
loss---the FaRM/RAMCloud trade~\citep{dragojevic2014farm}.
\textbf{No automated failover.} We validate the recovery mechanism and
inherit failure detection from the fabric, but leave primary promotion
(view change~\citep{oki1988vr,liskov2012vrrevisited}) to a production
layer; the recovery window of \S\ref{sec:membership} is the price of
passive replication's CPU savings.
\textbf{Shared-everything binaries.} Order-log-free replay covers
share-nothing servers; arbitrary shared-memory multithreading falls back
to the checkpoint-only path (\S\ref{sec:criu}).
\textbf{Residual non-determinism.} Two CPU instructions take no system
call and export no symbol: \texttt{RDRAND} (pinned per-consumer here) and
\texttt{RDTSC} (unused for externally visible state by our fifteen
applications). A fully general guarantee would trap both
(Appendix~\ref{app:libos}).
\textbf{The boundary of transparency.} The deepest limit is an
impossibility, not an engineering gap: no transparent system can un-send
an effect already delivered to an external party that will not cooperate
in deduplication---in essence the two-generals
obstacle~\citep{gray1978notes}. Output commit bounds the inconsistency
window; it cannot close it. The clean wins are therefore fabric-internal
services, idempotent interfaces, and self-contained leaf services; a
non-cooperating external peer forces output buffering at the boundary,
re-importing exactly the latency the conditions remove.

\section{Related Work}
\label{sec:related}

\paragraph{Transparent fault tolerance.} Whole-machine replication runs
unmodified workloads fault-tolerantly, from hypervisor-based
replication~\citep{bressoud1995hypervisor} through Remus~\citep{cully2008remus},
VMware FT~\citep{scales2010vmwareft}, COLO~\citep{dong2013colo}, and
Crane~\citep{cui2015crane}. Closest are LLFT~\citep{zhao2013llft}
(host-software total order, no order log, output released before backup
durability; \S\ref{sec:why}) and HyCoR~\citep{zhou2021hycor} (successor
RRC~\citep{zhou2022rrc}; container checkpoints plus short logged-order
replay windows). Relative to both, OneBarrier contributes the
four-condition factoring; its consequence---order \emph{and} durability in
the network, so the output hold coincides with a barrier the network
already crosses; and the breadth of evidence that \Determinism is closable
on real binaries.

\paragraph{Ordering and replication.} Total order via logical time
originates with \citet{lamport1978time}; atomic broadcast and virtual
synchrony made it a group-communication
primitive~\citep{birman1987isis,defago2004total}; the state-machine
approach~\citep{schneider1990smr} and consensus
(Paxos~\citep{lamport1998parttime}, Raft~\citep{ongaro2014raft},
Viewstamped Replication~\citep{oki1988vr}) built replication on it. A
recent line derives order from the network itself---NOPaxos~\citep{li2016nopaxos},
Eris~\citep{li2017eris}, Derecho~\citep{jha2019derecho}
(\S\ref{sec:why})---serving rewritten applications with
replicated execution. 1Pipe~\citep{li2021onepipe} provides
total-order \emph{communication} rather than replication; OneBarrier
repurposes it as a substrate for passive, transparent fault tolerance.
Calvin~\citep{thomson2012calvin} sequences database transactions up front
for deterministic execution---the database-layer analogue of
fabric-supplied order, obtained by rewriting.

\paragraph{Deterministic replay and record/replay.}
ReVirt~\citep{dunlap2002revirt}, rr~\citep{ocallahan2017rr}, and
Castor~\citep{mashtizadeh2017castor} record non-determinism for replay and
find message/shared-memory order to be the dominant cost---the cost
\Order removes at the source. Deterministic multithreading
(Kendo~\citep{olszewski2009kendo}, CoreDet~\citep{bergan2010coredet},
dThreads~\citep{liu2011dthreads}) and execute-then-verify
replication~\citep{kapritsos2012eve,guo2014rex} trade throughput for
thread-order determinism, the trade our sharding path avoids.

\paragraph{Checkpointing and the rewrite path.} CRIU~\citep{criu},
DMTCP~\citep{ansel2009dmtcp}, and BLCR~\citep{hargrove2006blcr} checkpoint
unmodified processes; microVM snapshots~\citep{agache2020firecracker,
ustiugov2021reap} and gVisor~\citep{gvisor} are the systems-level
analogues. Durable execution (Temporal~\citep{temporal},
DBOS~\citep{skiadopoulos2021dbos}, Restate~\citep{restate}) and exactly-once
streaming (Flink~\citep{carbone2015flink}) obtain the same guarantees by
rewriting. RIFL~\citep{lee2015rifl} supplies the exactly-once RPC
discipline our output suppression adapts.

\paragraph{Operating point and lineage.} The microsecond regime is set by
RDMA systems such as FaRM~\citep{dragojevic2014farm},
eRPC~\citep{kalia2019erpc}, and FastWake~\citep{li2023fastwake}; our
user-space interposition follows
SocksDirect~\citep{li2019socksdirect}, with kernel-bypass datapaths in the
Arrakis/Demikernel lineage~\citep{peter2014arrakis,zhang2021demikernel}.
The problems themselves are classical: output
commit~\citep{strom1985optimistic}, consistent
snapshots~\citep{chandy1985snapshots}, rollback-recovery at
large~\citep{elnozahy2002survey}, and
linearizability~\citep{herlihy1990linearizability}. Stateful
network-function recovery (FTMB~\citep{sherry2015ftmb},
Pico~\citep{rajagopalan2013pico}) is a per-domain instance of the same
replay problem.

\section{Conclusion}
\label{sec:conclusion}

Transparent fault tolerance did not fail because recovery is hard; it
failed because every system paid for ordering, snapshot coordination, and
output holding on networks that provided none of them. Stated as
conditions, the diagnosis becomes a design: a network that delivers in one
global order (\Order), confirms delivery at a commit barrier (\Barrier),
and replicates each message within that barrier (\Durability) eliminates
all three costs structurally, and a user-space shim closes the remaining
local non-determinism (\Determinism) of unmodified binaries at
$2$--$10\%$ overhead. On a
fabric that meets the conditions at microsecond scale, fifteen unmodified
applications---caches, web servers, brokers, databases, a network
function---recover byte-identically, crash injection confirms linearizable
histories and exactly-once effects, and the durable write that transparent
systems have always stacked on the critical path instead rides a barrier
the network was crossing anyway.
Fault tolerance, on such a network, is not a tax. It is a property.

\section*{Acknowledgements}
This paper began as a draft written during the author's internship at Microsoft Research in 2017.
Nine years later, with the help of Pine Copilot and Claude Code, the author finally brought it to completion.
The work was produced using Pine Copilot's voice-directed \emph{whisper coding}
workflow~\citep{pineai2026whispercoding}, in which the author specifies, discusses, and reviews
the work by voice while a coding agent---Claude Code with Claude Opus 4.8 and Claude Fable
5---carries out the planning, coding, experiments, and paper writing.
The author thanks BSQL Networking for hosting the NVIDIA RTX PRO 6000 GPU.

\bibliographystyle{plainnat}
\bibliography{references}

\begin{thebibliography}{59}
\providecommand{\natexlab}[1]{#1}
\providecommand{\url}[1]{\texttt{#1}}
\expandafter\ifx\csname urlstyle\endcsname\relax
  \providecommand{\doi}[1]{doi: #1}\else
  \providecommand{\doi}{doi: \begingroup \urlstyle{rm}\Url}\fi

\bibitem[Agache et~al.(2020)Agache, Brooker, Florescu, Iordache, Liguori,
  Neugebauer, Piwonka, and Popa]{agache2020firecracker}
Alexandru Agache, Marc Brooker, Andreea Florescu, Alexandra Iordache, Anthony
  Liguori, Rolf Neugebauer, Phil Piwonka, and Diana-Maria Popa.
\newblock Firecracker: Lightweight virtualization for serverless applications.
\newblock In \emph{Proceedings of the 17th USENIX Symposium on Networked
  Systems Design and Implementation (NSDI '20)}, pages 419--434. USENIX
  Association, 2020.

\bibitem[Ansel et~al.(2009)Ansel, Arya, and Cooperman]{ansel2009dmtcp}
Jason Ansel, Kapil Arya, and Gene Cooperman.
\newblock {DMTCP}: Transparent checkpointing for cluster computations and the
  desktop.
\newblock In \emph{Proceedings of the 23rd IEEE International Parallel and
  Distributed Processing Symposium (IPDPS '09)}. IEEE, 2009.
\newblock \doi{10.1109/IPDPS.2009.5161063}.

\bibitem[Bergan et~al.(2010)Bergan, Anderson, Devietti, Ceze, and
  Grossman]{bergan2010coredet}
Tom Bergan, Owen Anderson, Joseph Devietti, Luis Ceze, and Dan Grossman.
\newblock Coredet: A compiler and runtime system for deterministic
  multithreaded execution.
\newblock In \emph{Proceedings of the 15th International Conference on
  Architectural Support for Programming Languages and Operating Systems (ASPLOS
  '10)}, pages 53--64. ACM, 2010.

\bibitem[Birman and Joseph(1987)]{birman1987isis}
Kenneth~P. Birman and Thomas~A. Joseph.
\newblock Exploiting virtual synchrony in distributed systems.
\newblock In \emph{Proceedings of the 11th ACM Symposium on Operating Systems
  Principles (SOSP '87)}, pages 123--138. ACM, 1987.
\newblock \doi{10.1145/41457.37515}.

\bibitem[Bosshart et~al.(2014)Bosshart, Daly, Gibb, Izzard, McKeown, Rexford,
  Schlesinger, Talayco, Vahdat, Varghese, and Walker]{bosshart2014p4}
Pat Bosshart, Dan Daly, Glen Gibb, Martin Izzard, Nick McKeown, Jennifer
  Rexford, Cole Schlesinger, Dan Talayco, Amin Vahdat, George Varghese, and
  David Walker.
\newblock P4: Programming protocol-independent packet processors.
\newblock \emph{ACM SIGCOMM Computer Communication Review}, 44\penalty0
  (3):\penalty0 87--95, 2014.

\bibitem[Bressoud and Schneider(1995)]{bressoud1995hypervisor}
Thomas~C. Bressoud and Fred~B. Schneider.
\newblock Hypervisor-based fault tolerance.
\newblock In \emph{Proceedings of the Fifteenth ACM Symposium on Operating
  Systems Principles (SOSP '95)}, pages 1--11. ACM, 1995.
\newblock \doi{10.1145/224056.224058}.

\bibitem[Carbone et~al.(2015)Carbone, Katsifodimos, Ewen, Markl, Haridi, and
  Tzoumas]{carbone2015flink}
Paris Carbone, Asterios Katsifodimos, Stephan Ewen, Volker Markl, Seif Haridi,
  and Kostas Tzoumas.
\newblock Apache {Flink}: Stream and batch processing in a single engine.
\newblock \emph{Bulletin of the IEEE Computer Society Technical Committee on
  Data Engineering}, 38\penalty0 (4):\penalty0 28--38, 2015.

\bibitem[Chandy and Lamport(1985)]{chandy1985snapshots}
K.~Mani Chandy and Leslie Lamport.
\newblock Distributed snapshots: Determining global states of distributed
  systems.
\newblock \emph{ACM Transactions on Computer Systems}, 3\penalty0 (1):\penalty0
  63--75, 1985.
\newblock \doi{10.1145/214451.214456}.

\bibitem[Corbett et~al.(2012)Corbett, Dean, Epstein, Fikes, Frost, Furman,
  Ghemawat, Gubarev, Heiser, Hochschild, Hsieh, Kanthak, Kogan, Li, Lloyd,
  Melnik, Mwaura, Nagle, Quinlan, Rao, Rolig, Saito, Szymaniak, Taylor, Wang,
  and Woodford]{corbett2012spanner}
James~C. Corbett, Jeffrey Dean, Michael Epstein, Andrew Fikes, Christopher
  Frost, J.~J. Furman, Sanjay Ghemawat, Andrey Gubarev, Christopher Heiser,
  Peter Hochschild, Wilson Hsieh, Sebastian Kanthak, Eugene Kogan, Hongyi Li,
  Alexander Lloyd, Sergey Melnik, David Mwaura, David Nagle, Sean Quinlan,
  Rajesh Rao, Lindsay Rolig, Yasushi Saito, Michal Szymaniak, Christopher
  Taylor, Ruth Wang, and Dale Woodford.
\newblock Spanner: Google's globally-distributed database.
\newblock In \emph{Proceedings of the 10th USENIX Symposium on Operating
  Systems Design and Implementation (OSDI '12)}, pages 251--264. USENIX
  Association, 2012.

\bibitem[{CRIU Project}(2024)]{criu}
{CRIU Project}.
\newblock {CRIU} -- checkpoint/restore in userspace.
\newblock \url{https://criu.org/}, 2024.
\newblock Accessed 2026-06-23.

\bibitem[Cui et~al.(2015)Cui, Gu, Liu, Chen, and Yang]{cui2015crane}
Heming Cui, Rui Gu, Cheng Liu, Tianyu Chen, and Junfeng Yang.
\newblock Paxos made transparent.
\newblock In \emph{Proceedings of the 25th Symposium on Operating Systems
  Principles (SOSP '15)}. ACM, 2015.

\bibitem[Cully et~al.(2008)Cully, Lefebvre, Meyer, Feeley, Hutchinson, and
  Warfield]{cully2008remus}
Brendan Cully, Geoffrey Lefebvre, Dutch Meyer, Mike Feeley, Norm Hutchinson,
  and Andrew Warfield.
\newblock Remus: High availability via asynchronous virtual machine
  replication.
\newblock In \emph{Proceedings of the 5th USENIX Symposium on Networked Systems
  Design and Implementation (NSDI '08)}. USENIX Association, 2008.

\bibitem[D{\'e}fago et~al.(2004)D{\'e}fago, Schiper, and
  Urb{\'a}n]{defago2004total}
Xavier D{\'e}fago, Andr{\'e} Schiper, and P{\'e}ter Urb{\'a}n.
\newblock Total order broadcast and multicast algorithms: Taxonomy and survey.
\newblock \emph{ACM Computing Surveys}, 36\penalty0 (4):\penalty0 372--421,
  2004.
\newblock \doi{10.1145/1041680.1041682}.

\bibitem[Dong et~al.(2013)Dong, Ye, Jiang, Pratt, Ma, Li, and
  Guan]{dong2013colo}
YaoZu Dong, Wei Ye, YunHong Jiang, Ian Pratt, ShiQing Ma, Jian Li, and HaiBing
  Guan.
\newblock {COLO}: {COarse}-grained {LOck}-stepping virtual machines for
  non-stop service.
\newblock In \emph{Proceedings of the 4th Annual Symposium on Cloud Computing
  (SoCC '13)}. ACM, 2013.

\bibitem[Dragojevi{\'c} et~al.(2014)Dragojevi{\'c}, Narayanan, Castro, and
  Hodson]{dragojevic2014farm}
Aleksandar Dragojevi{\'c}, Dushyanth Narayanan, Miguel Castro, and Orion
  Hodson.
\newblock {FaRM}: Fast remote memory.
\newblock In \emph{Proceedings of the 11th USENIX Symposium on Networked
  Systems Design and Implementation (NSDI '14)}, pages 401--414. USENIX
  Association, 2014.

\bibitem[Dunlap et~al.(2002)Dunlap, King, Cinar, Basrai, and
  Chen]{dunlap2002revirt}
George~W. Dunlap, Samuel~T. King, Sukru Cinar, Murtaza~A. Basrai, and Peter~M.
  Chen.
\newblock {ReVirt}: Enabling intrusion analysis through virtual-machine logging
  and replay.
\newblock In \emph{Proceedings of the 5th Symposium on Operating Systems Design
  and Implementation (OSDI '02)}. USENIX Association, 2002.

\bibitem[Elnozahy et~al.(2002)Elnozahy, Alvisi, Wang, and
  Johnson]{elnozahy2002survey}
E.~N.~(Mootaz) Elnozahy, Lorenzo Alvisi, Yi-Min Wang, and David~B. Johnson.
\newblock A survey of rollback-recovery protocols in message-passing systems.
\newblock \emph{ACM Computing Surveys}, 34\penalty0 (3):\penalty0 375--408,
  2002.
\newblock \doi{10.1145/568522.568525}.

\bibitem[Fischer et~al.(1985)Fischer, Lynch, and
  Paterson]{fischer1985impossibility}
Michael~J. Fischer, Nancy~A. Lynch, and Michael~S. Paterson.
\newblock Impossibility of distributed consensus with one faulty process.
\newblock \emph{Journal of the ACM}, 32\penalty0 (2):\penalty0 374--382, 1985.
\newblock \doi{10.1145/3149.214121}.

\bibitem[Geng et~al.(2018)Geng, Liu, Yin, Naik, Prabhakar, Rosenblum, and
  Vahdat]{geng2018huygens}
Yilong Geng, Shiyu Liu, Zi~Yin, Ashish Naik, Balaji Prabhakar, Mendel
  Rosenblum, and Amin Vahdat.
\newblock Exploiting a natural network effect for scalable, fine-grained clock
  synchronization.
\newblock In \emph{Proceedings of the 15th USENIX Symposium on Networked
  Systems Design and Implementation (NSDI '18)}, pages 81--94. USENIX
  Association, 2018.

\bibitem[{Google}(2018)]{gvisor}
{Google}.
\newblock {gVisor}: The container security platform.
\newblock \url{https://gvisor.dev/}, 2018.
\newblock Accessed 2026-06-23; source: \url{https://github.com/google/gvisor}.

\bibitem[Gray(1978)]{gray1978notes}
Jim Gray.
\newblock Notes on data base operating systems.
\newblock In \emph{Operating Systems: An Advanced Course}, volume~60 of
  \emph{Lecture Notes in Computer Science}, pages 393--481. Springer, 1978.

\bibitem[Guo et~al.(2014)Guo, Hong, Yang, Zhou, Zhou, and Zhuang]{guo2014rex}
Zhenyu Guo, Chuntao Hong, Mao Yang, Dong Zhou, Lidong Zhou, and Li~Zhuang.
\newblock Rex: Replication at the speed of multi-core.
\newblock In \emph{Proceedings of the Ninth European Conference on Computer
  Systems (EuroSys '14)}. ACM, 2014.

\bibitem[Hargrove and Duell(2006)]{hargrove2006blcr}
Paul~H. Hargrove and Jason~C. Duell.
\newblock Berkeley lab checkpoint/restart ({BLCR}) for {Linux} clusters.
\newblock In \emph{Journal of Physics: Conference Series (SciDAC 2006)},
  volume~46, pages 494--499. IOP Publishing, 2006.
\newblock \doi{10.1088/1742-6596/46/1/067}.

\bibitem[Herlihy and Wing(1990)]{herlihy1990linearizability}
Maurice~P. Herlihy and Jeannette~M. Wing.
\newblock Linearizability: A correctness condition for concurrent objects.
\newblock \emph{ACM Transactions on Programming Languages and Systems},
  12\penalty0 (3):\penalty0 463--492, 1990.
\newblock \doi{10.1145/78969.78972}.

\bibitem[Jha et~al.(2019)Jha, Behrens, Gkountouvas, Milano, Song, Tremel,
  Renesse, Zink, and Birman]{jha2019derecho}
Sagar Jha, Jonathan Behrens, Theo Gkountouvas, Matthew Milano, Weijia Song,
  Edward Tremel, Robbert~Van Renesse, Sydney Zink, and Kenneth~P. Birman.
\newblock Derecho: Fast state machine replication for cloud services.
\newblock \emph{ACM Transactions on Computer Systems}, 36\penalty0 (2), 2019.

\bibitem[Kalia et~al.(2019)Kalia, Kaminsky, and Andersen]{kalia2019erpc}
Anuj Kalia, Michael Kaminsky, and David~G. Andersen.
\newblock Datacenter {RPCs} can be general and fast.
\newblock In \emph{Proceedings of the 16th USENIX Symposium on Networked
  Systems Design and Implementation (NSDI '19)}, pages 1--16. USENIX
  Association, 2019.

\bibitem[Kapritsos et~al.(2012)Kapritsos, Wang, Qu{\'e}ma, Clement, Alvisi, and
  Dahlin]{kapritsos2012eve}
Manos Kapritsos, Yang Wang, Vivien Qu{\'e}ma, Allen Clement, Lorenzo Alvisi,
  and Mike Dahlin.
\newblock All about {Eve}: Execute-verify replication for multi-core servers.
\newblock In \emph{Proceedings of the 10th USENIX Symposium on Operating
  Systems Design and Implementation (OSDI '12)}. USENIX Association, 2012.

\bibitem[Lamport(1978)]{lamport1978time}
Leslie Lamport.
\newblock Time, clocks, and the ordering of events in a distributed system.
\newblock \emph{Communications of the ACM}, 21\penalty0 (7):\penalty0 558--565,
  1978.
\newblock \doi{10.1145/359545.359563}.

\bibitem[Lamport(1998)]{lamport1998parttime}
Leslie Lamport.
\newblock The part-time parliament.
\newblock \emph{ACM Transactions on Computer Systems}, 16\penalty0
  (2):\penalty0 133--169, 1998.

\bibitem[Lee et~al.(2015)Lee, Park, Kejriwal, Matsushita, and
  Ousterhout]{lee2015rifl}
Collin Lee, Seo~Jin Park, Ankita Kejriwal, Satoshi Matsushita, and John
  Ousterhout.
\newblock Implementing linearizability at large scale and low latency.
\newblock In \emph{Proceedings of the 25th Symposium on Operating Systems
  Principles (SOSP '15)}, pages 71--86. ACM, 2015.
\newblock \doi{10.1145/2815400.2815416}.

\bibitem[Li et~al.(2016{\natexlab{a}})Li, Tan, Luo, Peng, Luo, Xu, Xiong,
  Cheng, and Chen]{li2016clicknp}
Bojie Li, Kun Tan, Layong~(Larry) Luo, Yanqing Peng, Renqian Luo, Ningyi Xu,
  Yongqiang Xiong, Peng Cheng, and Enhong Chen.
\newblock {ClickNP}: Highly flexible and high-performance network processing
  with reconfigurable hardware.
\newblock In \emph{Proceedings of the 2016 ACM SIGCOMM Conference (SIGCOMM
  '16)}, pages 1--14. ACM, 2016{\natexlab{a}}.
\newblock \doi{10.1145/2934872.2934897}.

\bibitem[Li et~al.(2019)Li, Cui, Wang, Bai, and Zhang]{li2019socksdirect}
Bojie Li, Tianyi Cui, Zibo Wang, Wei Bai, and Lintao Zhang.
\newblock {SocksDirect}: Datacenter sockets can be fast and compatible.
\newblock In \emph{Proceedings of the ACM Special Interest Group on Data
  Communication (SIGCOMM '19)}. ACM, 2019.

\bibitem[Li et~al.(2021)Li, Zuo, Bai, and Zhang]{li2021onepipe}
Bojie Li, Gefei Zuo, Wei Bai, and Lintao Zhang.
\newblock {1Pipe}: Scalable total order communication in data center networks.
\newblock In \emph{Proceedings of the 2021 ACM SIGCOMM Conference (SIGCOMM
  '21)}. ACM, 2021.

\bibitem[Li et~al.(2023)Li, Xiang, Wang, Ruan, Zhou, and Tan]{li2023fastwake}
Bojie Li, Zhihao Xiang, Xiaoliang Wang, Han Ruan, Jingbin Zhou, and Kun Tan.
\newblock {FastWake}: Revisiting host network stack for interrupt-mode {RDMA}.
\newblock In \emph{Proceedings of the 7th Asia-Pacific Workshop on Networking
  (APNet '23)}, 2023.

\bibitem[Li et~al.(2016{\natexlab{b}})Li, Michael, Sharma, Szekeres, and
  Ports]{li2016nopaxos}
Jialin Li, Ellis Michael, Naveen~Kr. Sharma, Adriana Szekeres, and Dan R.~K.
  Ports.
\newblock Just say {NO} to paxos overhead: Replacing consensus with network
  ordering.
\newblock In \emph{Proceedings of the 12th USENIX Symposium on Operating
  Systems Design and Implementation (OSDI '16)}, pages 467--483. USENIX
  Association, 2016{\natexlab{b}}.

\bibitem[Li et~al.(2017)Li, Michael, and Ports]{li2017eris}
Jialin Li, Ellis Michael, and Dan R.~K. Ports.
\newblock Eris: Coordination-free consistent transactions using in-network
  concurrency control.
\newblock In \emph{Proceedings of the 26th ACM Symposium on Operating Systems
  Principles (SOSP '17)}, pages 104--120. ACM, 2017.

\bibitem[Liskov and Cowling(2012)]{liskov2012vrrevisited}
Barbara Liskov and James Cowling.
\newblock Viewstamped replication revisited.
\newblock Technical Report MIT-CSAIL-TR-2012-021, MIT Computer Science and
  Artificial Intelligence Laboratory (CSAIL), 2012.

\bibitem[Liu et~al.(2011)Liu, Curtsinger, and Berger]{liu2011dthreads}
Tongping Liu, Charlie Curtsinger, and Emery~D. Berger.
\newblock Dthreads: Efficient deterministic multithreading.
\newblock In \emph{Proceedings of the 23rd ACM Symposium on Operating Systems
  Principles (SOSP '11)}, pages 327--336. ACM, 2011.

\bibitem[Mashtizadeh et~al.(2017)Mashtizadeh, Garfinkel, Terei, Mazi{\`e}res,
  and Rosenblum]{mashtizadeh2017castor}
Ali~Jos{\'e} Mashtizadeh, Tal Garfinkel, David Terei, David Mazi{\`e}res, and
  Mendel Rosenblum.
\newblock Towards practical default-on multi-core record/replay.
\newblock In \emph{Proceedings of the 22nd International Conference on
  Architectural Support for Programming Languages and Operating Systems (ASPLOS
  '17)}. ACM, 2017.

\bibitem[O'Callahan et~al.(2017)O'Callahan, Jones, Froyd, Huey, Noll, and
  Partush]{ocallahan2017rr}
Robert O'Callahan, Chris Jones, Nathan Froyd, Kyle Huey, Albert Noll, and
  Nimrod Partush.
\newblock Engineering record and replay for deployability.
\newblock In \emph{Proceedings of the 2017 USENIX Annual Technical Conference
  (USENIX ATC '17)}. USENIX Association, 2017.

\bibitem[Oki and Liskov(1988)]{oki1988vr}
Brian~M. Oki and Barbara~H. Liskov.
\newblock Viewstamped replication: A new primary copy method to support
  highly-available distributed systems.
\newblock In \emph{Proceedings of the Seventh Annual ACM Symposium on
  Principles of Distributed Computing (PODC '88)}, pages 8--17. ACM, 1988.

\bibitem[Olszewski et~al.(2009)Olszewski, Ansel, and
  Amarasinghe]{olszewski2009kendo}
Marek Olszewski, Jason Ansel, and Saman Amarasinghe.
\newblock Kendo: Efficient deterministic multithreading in software.
\newblock In \emph{Proceedings of the 14th International Conference on
  Architectural Support for Programming Languages and Operating Systems (ASPLOS
  '09)}, pages 97--108. ACM, 2009.

\bibitem[Ongaro and Ousterhout(2014)]{ongaro2014raft}
Diego Ongaro and John Ousterhout.
\newblock In search of an understandable consensus algorithm.
\newblock In \emph{Proceedings of the 2014 USENIX Annual Technical Conference
  (USENIX ATC '14)}, pages 305--319. USENIX Association, 2014.

\bibitem[Peter et~al.(2014)Peter, Li, Zhang, Ports, Woos, Krishnamurthy,
  Anderson, and Roscoe]{peter2014arrakis}
Simon Peter, Jialin Li, Irene Zhang, Dan R.~K. Ports, Doug Woos, Arvind
  Krishnamurthy, Thomas Anderson, and Timothy Roscoe.
\newblock Arrakis: The operating system is the control plane.
\newblock In \emph{Proceedings of the 11th USENIX Symposium on Operating
  Systems Design and Implementation (OSDI '14)}. USENIX Association, 2014.

\bibitem[{Pine AI}(2026)]{pineai2026whispercoding}
{Pine AI}.
\newblock {Pine AI}: The most natural human-computer interface is your voice.
\newblock
  \url{https://www.19pine.ai/blog/pine-ai-the-most-natural-human-computer-interface-is-your-voice},
  2026.
\newblock Blog post; accessed 2026-07-02.

\bibitem[Rajagopalan et~al.(2013)Rajagopalan, Williams, Jamjoom, and
  Warfield]{rajagopalan2013pico}
Shriram Rajagopalan, Dan Williams, Hani Jamjoom, and Andrew Warfield.
\newblock {Pico Replication}: A high availability framework for middleboxes.
\newblock In \emph{Proceedings of the 4th Annual Symposium on Cloud Computing
  (SoCC '13)}, pages 1--15. ACM, 2013.
\newblock \doi{10.1145/2523616.2523635}.

\bibitem[{Restate}(2024)]{restate}
{Restate}.
\newblock Restate: Durable execution platform.
\newblock \url{https://docs.restate.dev/}, 2024.
\newblock Industry durable-execution system; accessed 2026-06-23.

\bibitem[Scales et~al.(2010)Scales, Nelson, and
  Venkitachalam]{scales2010vmwareft}
Daniel~J. Scales, Mike Nelson, and Ganesh Venkitachalam.
\newblock The design of a practical system for fault-tolerant virtual machines.
\newblock \emph{ACM SIGOPS Operating Systems Review}, 44\penalty0 (4):\penalty0
  30--39, 2010.

\bibitem[Schneider(1990)]{schneider1990smr}
Fred~B. Schneider.
\newblock Implementing fault-tolerant services using the state machine
  approach: A tutorial.
\newblock \emph{ACM Computing Surveys}, 22\penalty0 (4):\penalty0 299--319,
  1990.

\bibitem[Sherry et~al.(2015)Sherry, Gao, Basu, Panda, Krishnamurthy, Maciocco,
  Manesh, Martins, Ratnasamy, Rizzo, and Shenker]{sherry2015ftmb}
Justine Sherry, Peter~Xiang Gao, Soumya Basu, Aurojit Panda, Arvind
  Krishnamurthy, Christian Maciocco, Maziar Manesh, Jo{\~a}o Martins, Sylvia
  Ratnasamy, Luigi Rizzo, and Scott Shenker.
\newblock Rollback-recovery for middleboxes.
\newblock In \emph{Proceedings of the 2015 ACM SIGCOMM Conference (SIGCOMM
  '15)}, pages 227--240. ACM, 2015.
\newblock \doi{10.1145/2785956.2787501}.

\bibitem[Skiadopoulos et~al.(2021)Skiadopoulos, Li, Kraft, Kaffes, Hong,
  Mathew, Bestor, Cafarella, Gadepally, Graefe, Kepner, Kozyrakis, Kraska,
  Stonebraker, Suresh, and Zaharia]{skiadopoulos2021dbos}
Athinagoras Skiadopoulos, Qian Li, Peter Kraft, Kostis Kaffes, Daniel Hong,
  Shana Mathew, David Bestor, Michael Cafarella, Vijay Gadepally, Goetz Graefe,
  Jeremy Kepner, Christos Kozyrakis, Tim Kraska, Michael Stonebraker, Lalith
  Suresh, and Matei Zaharia.
\newblock {DBOS}: A {DBMS}-oriented operating system.
\newblock \emph{Proceedings of the VLDB Endowment}, 15\penalty0 (1):\penalty0
  21--30, 2021.

\bibitem[Strom and Yemini(1985)]{strom1985optimistic}
Robert~E. Strom and Shaula Yemini.
\newblock Optimistic recovery in distributed systems.
\newblock \emph{ACM Transactions on Computer Systems}, 3\penalty0 (3):\penalty0
  204--226, 1985.
\newblock \doi{10.1145/3959.3962}.

\bibitem[{Temporal Technologies}(2024)]{temporal}
{Temporal Technologies}.
\newblock Temporal: Durable execution platform.
\newblock \url{https://docs.temporal.io/}, 2024.
\newblock Industry durable-execution system; accessed 2026-06-23.

\bibitem[Thomson et~al.(2012)Thomson, Diamond, Weng, Ren, Shao, and
  Abadi]{thomson2012calvin}
Alexander Thomson, Thaddeus Diamond, Shu-Chun Weng, Kun Ren, Philip Shao, and
  Daniel~J. Abadi.
\newblock Calvin: Fast distributed transactions for partitioned database
  systems.
\newblock In \emph{Proceedings of the 2012 ACM SIGMOD International Conference
  on Management of Data (SIGMOD '12)}, pages 1--12. ACM, 2012.
\newblock \doi{10.1145/2213836.2213838}.

\bibitem[Ustiugov et~al.(2021)Ustiugov, Petrov, Kogias, Bugnion, and
  Grot]{ustiugov2021reap}
Dmitrii Ustiugov, Plamen Petrov, Marios Kogias, Edouard Bugnion, and Boris
  Grot.
\newblock Benchmarking, analysis, and optimization of serverless function
  snapshots.
\newblock In \emph{Proceedings of the 26th International Conference on
  Architectural Support for Programming Languages and Operating Systems (ASPLOS
  '21)}, pages 559--572. ACM, 2021.
\newblock \doi{10.1145/3445814.3446714}.

\bibitem[Zhang et~al.(2021)Zhang, Raybuck, Patel, Olynyk, Nelson, Leija,
  Martinez, Liu, Simpson, Jayakar, Penna, Demoulin, Choudhury, and
  Badam]{zhang2021demikernel}
Irene Zhang, Amanda Raybuck, Pratyush Patel, Kirk Olynyk, Jacob Nelson, Omar
  S.~Navarro Leija, Ashlie Martinez, Jing Liu, Anna~Kornfeld Simpson, Sujay
  Jayakar, Pedro~Henrique Penna, Max Demoulin, Piali Choudhury, and Anirudh
  Badam.
\newblock The {Demikernel} datapath {OS} architecture for microsecond-scale
  datacenter systems.
\newblock In \emph{Proceedings of the ACM SIGOPS 28th Symposium on Operating
  Systems Principles (SOSP '21)}. ACM, 2021.

\bibitem[Zhao et~al.(2013)Zhao, Melliar-Smith, and Moser]{zhao2013llft}
Wenbing Zhao, P.~M. Melliar-Smith, and L.~E. Moser.
\newblock Low latency fault tolerance system.
\newblock \emph{The Computer Journal}, 56\penalty0 (6):\penalty0 716--740,
  2013.

\bibitem[Zhou and Tamir(2021)]{zhou2021hycor}
Diyu Zhou and Yuval Tamir.
\newblock {HyCoR}: Fault-tolerant replicated containers based on checkpoint and
  replay.
\newblock \emph{CoRR}, abs/2101.09584, 2021.
\newblock arXiv preprint arXiv:2101.09584.

\bibitem[Zhou and Tamir(2022)]{zhou2022rrc}
Diyu Zhou and Yuval Tamir.
\newblock {RRC}: Responsive replicated containers.
\newblock In \emph{2022 USENIX Annual Technical Conference (USENIX ATC '22)}.
  USENIX Association, 2022.

\end{thebibliography}

\appendix

\section{Machine-Checked Specifications}
\label{app:tla}
The two protocols of \S\ref{sec:impl} are specified in TLA+ and checked
with the TLC model checker. The first specification models the fabric's
timestamping, barrier propagation, and delivery gate, and checks
\emph{total order}: every receiver delivers messages in one global order
consistent with the timestamps. TLC explores $3.5\times10^{6}$ reachable
states with no violation. The second models the replay
engine---the durable ordered log, snapshots, the per-client high-water
marks, and crash/recover transitions---and checks two invariants under
arbitrary crash and recovery interleavings: \textsc{ExactlyOnce} (each
acknowledged client write is applied to recovered state exactly once) and
\textsc{NoLostAck} (every acknowledged write is present after recovery).
Both specifications ship with the open-source artifact.

\section{The Discrete-Event Model and Its Calibration}
\label{app:model}

This appendix documents how the magnitude of the ride-versus-stack
result (\S\ref{sec:eval-overlap}) at the microsecond operating point is
modeled, calibrated, and corroborated.

\paragraph{Model.} Each request traverses four stages---order assignment,
the reliable-delivery barrier, the replication scatter (present only under
fault tolerance), and commit/release---with service times taken from
1Pipe's published measurements (Table~\ref{tab:model}) and open Poisson
arrivals swept to saturation. The replication scatter shares the message's
total-order position, so it falls within the barrier stage; the
out-of-regime comparison point instead places a serial $100\,\mu s$
storage write after the barrier.

\begin{table}[ht]\centering\small
\begin{tabular}{lr}
\toprule
model parameter (from 1Pipe~\citep{li2021onepipe}) & value \\
\midrule
end-host round-trip time & $1$--$2\,\mu s$ \\
order assignment & $0.5$\,RTT \\
best-effort delivery & $\sim$$10\,\mu s$ \\
reliable-delivery barrier & $\sim$$21\,\mu s$ ($1.5$\,RTT $+$ barrier wait) \\
replication scatter (FT only) & $\sim$$1.5\,\mu s$ ($1$\,RTT) \\
serial stable-storage write (out of regime) & $\sim$$100\,\mu s$ \\
per-host service cap & $80$\,M msg/s \\
\bottomrule
\end{tabular}
\caption{Discrete-event model parameters.}
\label{tab:model}
\end{table}

\paragraph{Calibration against the real engine.} Fed the engine's measured
loopback latencies instead of the RDMA parameters, the model reproduces
the in-barrier commit latency at matched load to within $\sim$$0.1\%$, and
predicts that the serial-fsync tier saturates at $\approx$$340$\,ops/s;
the real engine holds near its latency floor below $\sim$$300$\,ops/s and
diverges past $\sim$$370$\,ops/s, bracketing the predicted knee on real
software. Only the absolute RDMA latencies are substituted when the model
is run at the microsecond point. (At the loopback point the serial tier is
bound by the \emph{measured} $\sim$$3$\,ms fsync rather than the
$100\,\mu s$ table value, which is why its collapse there is
$\sim$$6000\times$ rather than $200\times$.)

\begin{figure}[ht]
\centering
\includegraphics[width=0.6\linewidth]{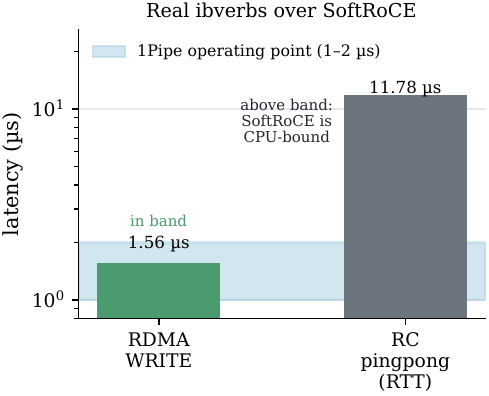}
\caption{Real RDMA transfers over SoftRoCE. The one-sided write sits
inside the modeled $1$--$2\,\mu s$ band; the round trip is CPU-bound in
software and lands above it.}
\label{fig:softroce}
\end{figure}

\paragraph{Real RDMA transfers.} Over SoftRoCE (a software implementation
of the RDMA protocol on ordinary Ethernet), a one-sided
\texttt{RDMA\_WRITE} completes in $\sim$$1.5\,\mu s$, inside the
$1$--$2\,\mu s$ band the model assumes for the replication write; the full
round trip measures $\sim$$11.8\,\mu s$ because SoftRoCE's receive path is
CPU-bound where hardware is not (Figure~\ref{fig:softroce}). The transfers
corroborate the one-sided-write latency, not the end-to-end pipeline,
which only a hardware testbed would settle.

\begin{figure}[ht]
\centering
\includegraphics[width=0.6\linewidth]{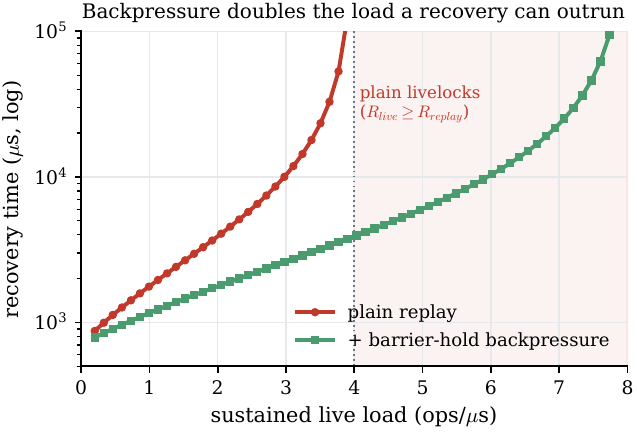}
\caption{Recovery time versus sustained live load (catch-up model). Plain
replay livelocks once the live rate reaches replay capacity; barrier-hold
backpressure roughly doubles the load a recovery can outrun.}
\label{fig:recovery_load}
\end{figure}

\paragraph{Recovery under sustained load.} In a closed-form catch-up model
parameterized by the engine's measured replay rate, plain replay livelocks
once the live rate reaches replay capacity, while the fabric's barrier
hold, used as backpressure on senders, roughly doubles the live load a
recovery can outrun while keeping recovery time bounded
(Figure~\ref{fig:recovery_load}). The real engine's recovery under
concurrent client load is exercised separately by the crash-injection test
of \S\ref{sec:eval-correct}; a hardware saturation sweep remains future
work.

\section{Determinism-Shim Details}
\label{app:libos}

\paragraph{Virtual clock.} Listing~\ref{lst:vclock} shows the interceptor.
The unit of advancement is one delivered message, not one TCP segment: the
fabric hands the server one message per \texttt{recv}, so record and
replay consume identical read sequences. One residual: time advances only
at input events, so an idle server's clock is frozen between requests and
a TTL due during an idle gap is observed at the next input---the value is
correct, the observation deferred. The vDSO fast path is not an escape
hatch, because \texttt{LD\_PRELOAD} overrides the exported symbols.

\begin{lstlisting}[style=cstyle,caption={The virtual clock. At each input
event, time advances by the \emph{real} inter-arrival gap, logged live and
replayed on recovery. Over the fabric, \texttt{delta} is derived from the
message's total-order timestamp rather than measured.},label={lst:vclock},float=ht,
basicstyle=\ttfamily\footnotesize]
ssize_t recv(int fd, void *b, size_t n, int fl) {
  ssize_t r = real_recv(fd, b, n, fl);
  if (r > 0 && is_conn(fd)) {           // a deterministic input event:
    long long delta = recording         //   wall-clock delta since last input,
        ? log_delta(now() - last)       //   measured + logged live, or
        : replay_delta();               //   replayed from the log on recovery
    atomic_fetch_add(&ticks, delta);    //   advance virtual time by it
  }
  return r;
}
int clock_gettime(clockid_t c, struct timespec *t) {
  long long v = base_ns + atomic_load(&ticks);  // base persisted at record
  t->tv_sec = v / 1000000000; t->tv_nsec = v % 1000000000;
  return 0;                        // same inputs+deltas -> same time
}
\end{lstlisting}

\paragraph{The \texttt{RDRAND}/\texttt{RDTSC} holes.} Time and randomness
are virtualized at the library, system-call, and device-file surfaces, but
two user-space CPU instructions take no system call and export no symbol.
\texttt{RDRAND} feeds OpenSSL (masked via \texttt{OPENSSL\_ia32cap}) and
V8's \texttt{Math.random} seed, which we pin with a recorded
\texttt{--random-seed} launch flag. \texttt{RDTSC} is used by none of our
fifteen applications for externally visible state but is not yet generally
pinned. A fully general guarantee would trap both---\texttt{RDTSC} via
\texttt{prctl(PR\_SET\_TSC)} and \texttt{RDRAND} via the instruction-trap
path---and emulate them from the virtual clock and seed.

\paragraph{Engineering realities.} {\sloppy Making ``unmodified'' credible
meant handling the awkward edges of interposition: the early-initialization
trap of \texttt{LD\_PRELOAD}; glibc symbol versioning
(\texttt{pthread\_cond\_wait} must be bound with \texttt{dlvsym} to
\texttt{GLIBC\_2.3.2}, since plain \texttt{dlsym} returns an old
compatibility shim and hangs threaded servers); a seccomp filter whose
supervisor lives in the very process it traps; and disabling Memcached's
timer-driven maintenance threads. These edges are the substance of the
per-consumer boundary of \S\ref{sec:rng}: library-level interposition
alone misses the raw-syscall and internal-call paths real servers use.\par}

\section{The Extension Study}
\label{app:ext}

This appendix expands \S\ref{sec:eval-recovery}'s ten extension
applications. The fit test's fourth precondition---bounded output---is an
output-commit requirement: suppression keys each externalized effect by a
per-client high-water mark, so an application whose response is an
unbounded stream has no sequence point at which an effect is acknowledged
and replay-suppressible. Request/response and discrete-message workloads
qualify; an infinite server-push stream does not.

\begin{figure}[ht]
\centering
\includegraphics[width=0.5\linewidth]{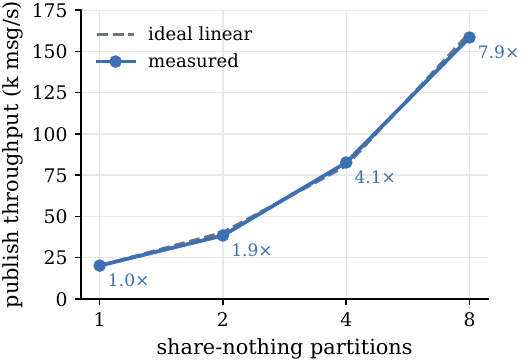}
\caption{Kafka-style partitions: $N$ share-nothing single-thread
partitions scale publish throughput near-linearly, reaching $7.9\times$ at
$8$ partitions.}
\label{fig:partscale}
\end{figure}

\paragraph{Message brokers.} A broker's job---total order plus durable
delivery---is what the fabric already provides. A stock
\texttt{redis-server} used as a stream broker recovers entry IDs and
consumer offsets byte-identically, because a Redis stream auto-ID derives
from the server clock the virtual clock determinizes. The Kafka partition
model \emph{is} share-nothing sharding---$N$ single-threaded partitions,
each with its own log---so a killed partition recovers byte-identically
with offsets preserved while publish throughput scales near-linearly
(Figure~\ref{fig:partscale}). Mosquitto carries the result to MQTT,
recovering its retained-message store by replaying the publish stream.

\paragraph{Databases.} PostgreSQL recovers only via CRIU
(\S\ref{sec:criu}) because it is shared-everything. SQLite is
single-threaded, so a stock \texttt{python3} process using the standard
\texttt{sqlite3} module recovers by deterministic \emph{replay}: rows
derived from SQLite's own time (\texttt{strftime}) and randomness
(\texttt{random()}, seeded from \texttt{/dev/urandom}) are byte-identical
across a crash where a control differs in both. The checkpoint path covers
the rest: PostgreSQL's process tree and a multi-threaded MySQL/MariaDB
both checkpoint-restore byte-identically with the server live.

\paragraph{A stateful network function.} A network function is a
deterministic packet-processing state machine, so its flow state is
recoverable by replaying the ordered packet log---the stateful-middlebox
problem of FTMB and Pico~\citep{sherry2015ftmb,rajagopalan2013pico}. A
software Click pipeline of ClickNP-style elements~\citep{li2016clicknp} (a
flow cache plus an L4 load balancer) recovers each flow's backend affinity
and connection-tracking timestamps byte-identically, where a control's
timestamps differ; a stateless restart cannot reconstruct this at all.

\paragraph{A stateful microservice.} Durable-execution frameworks rewrite
a service to externalize its state and non-deterministic effects.
OneBarrier gives an unmodified order/checkout microservice the same
guarantees transparently: across a crash, the order book recovers
exactly-once and byte-identically---including the random order IDs and
timestamps a rewrite would externalize by hand. The honest boundary
(\S\ref{sec:limits}): a service that calls other, non-cooperating services
cannot un-send those calls, so the clean wins are leaf services.

\paragraph{Infrastructure daemons, and a UDP lesson.} lighttpd and HAProxy
recover their \texttt{Date} headers byte-identically as Nginx does, and
dnsmasq's cached-record TTLs recover exactly under the virtual clock while
a control tracks real time. dnsmasq also exposed a boundary worth
recording: DNS runs over UDP, and the virtual clock originally ticked only
on TCP reads. A small extension ticking on
\texttt{recvfrom}/\texttt{recvmsg} makes single-process UDP servers
deterministic---and since dnsmasq's TCP path forks per query, UDP turned
out to be the \emph{deterministic} path, the reverse of the usual
assumption. The determinism unit is the message, not the connection.

\paragraph{Cross-shard atomicity.} The natural objection to sharding is
multi-key atomicity, which the fabric answers by construction: a scatter
delivers a group of messages to different destinations at one total-order
position, atomically---a multi-shard write without application-level
two-phase commit. We note this as a design consequence, not a measured
result.

\end{document}